\documentclass[12pt]{article}
\pdfoutput=1

\usepackage{amsfonts}
\usepackage{hyperref}
\newlength{\abstractwidth}
\usepackage{epsf}
\usepackage{color}
\usepackage{graphicx}

\renewcommand{\thefootnote}{\fnsymbol{footnote}}
\renewcommand{\thanks}[1]{\footnote{#1}}
\newcommand{\starttext}{
\setcounter{footnote}{0}
\renewcommand{\thefootnote}{\arabic{footnote}}}

\newcommand{\bea}{\begin{eqnarray}}
\newcommand{\eea}{\end{eqnarray}}
\newcommand{\ee}{\end{equation}}
\newcommand{\be}{\begin{equation}}

\def\cD{{\cal D}}

\def\cM{{\cal M}}
\def\cN{{\cal N}}
\def\cO{{\cal O}}

\def\bC{{\bf C}}

\def\bR{{\bf R}}
\def\bZ{{\bf Z}}
\def\Re{{\rm Re}}
\def\Im{{\rm Im}}

\def\det{{\rm det}}

\def\half{ {1\over 2}}
\def\p{\partial}

\def\a{\alpha}
\def\b{\beta}

\def\ep{\varepsilon}

\def\g{\gamma}
\def\k{\kappa}

\def\f{\varphi}

\def\ch{{\rm \, ch }}
\def\sh{{\rm \, sh }}
\def\th{{\rm \, th }}
\def\tg{{\rm \, tg  }}
\def\cn{{\rm \, cn }}
\def\sn{{\rm \, sn }}
\def\dn{{\rm \, dn }}

\def\bu{{\bar u}}

\def\bw{{\bar w}}
\def\bz{{\bar z}}

\def\no{\nonumber}
\def\sm{\smallskip}

\long\def\symbolfootnote[#1]#2{\begingroup%
\def\thefootnote{\fnsymbol{footnote}}\footnote[#1]{#2}\endgroup}

\begin{document}
\starttext
\setcounter{footnote}{0}

\begin{flushright}
UCLA/09/TEP/44 \\
CPHT-RR042.0509 \\
2 June  2009
\end{flushright}

\bigskip

\begin{center}

{\Large \bf Exact Half-BPS Flux Solutions in
M-theory III }\symbolfootnote[2]{\noindent This work was
supported in part by NSF grant PHY-07-57702.}

\medskip

{\large \bf Existence and rigidity of global solutions  asymptotic to $AdS_4 \times S^7$}

\medskip

\vskip .4in

{\large  Eric D'Hoker$^{a}$, John Estes$^{b}$, Michael Gutperle$^{a}$,
and  Darya Krym$^{a}$}

\vskip .2in

{$\ ^{a}$ \sl Department of Physics and Astronomy }\\
{\sl University of California, Los Angeles, CA 90095, USA}\\
{\tt \small dhoker@physics.ucla.edu; gutperle@physics.ucla.edu;
dk320@physics.ucla.edu}

\vskip .2in

{$\ ^{b}$\sl  Centre de Physique Th«eorique, Ecole Polytechnique, CNRS}\\
{\sl 91128 Palaiseau, France}\\
{\tt  \small johnaldonestes@gmail.com}
\end{center}

\vskip .2in

\begin{abstract}

\vskip 0.1in

The BPS equations in M-theory for solutions with 16 residual supersymmetries,
$SO(2,2)\times SO(4)\times SO(4)$ symmetry, and  $AdS_4 \times S^7$
asymptotics, were reduced in [arXiv:0806.0605] to a linear first order partial
differential equation on a Riemann surface with boundary, subject to a non-trivial
quadratic constraint. In the present paper, suitable regularity and boundary
conditions are imposed for the existence of global solutions. We seek regular solutions
with multiple distinct asymptotic $AdS_4 \times S^7$ regions, but find that, remarkably,
such solutions invariably reduce to multiple covers of the M-Janus
solution found by the authors  in [arXiv:0904.3313], suggesting rigidity of the half-BPS
M-Janus solution. In particular, we prove analytically that no other smooth
deformations away from the M-Janus solution exist, as such deformations
invariably violate the quadratic constraint. These rigidity results are contrasted to the
existence of half-BPS solutions with non-trivial 4-form fluxes and charges
asymptotic to $AdS_7 \times S^4$. The results are related to the possibility of
M2-branes to end on M5-branes, but the impossibility of M5-branes to end
on M2-branes, and to the non-existence of half-BPS solutions with simultaneous
$AdS_4 \times S^7$ and $AdS_7 \times S^4$ asymptotic regions.

\end{abstract}

\newpage


\newpage

\baselineskip=16pt
\setcounter{equation}{0}
\setcounter{footnote}{0}

\section{Introduction}
\setcounter{equation}{0}

One of the most important  realizations of the  AdS/CFT correspondence
\cite{Maldacena:1997re,Gubser:1998bc,Witten:1998qj} (for reviews,
see \cite{Aharony:1999ti,D'Hoker:2002aw}) in M-theory is the duality
of  the $AdS_{4}\times S^{7}$ vacuum and the 3-dimensional CFT which is
obtained by a decoupling limit of the M2-brane world-volume theory
\cite{Aharony:1998rm,Minwalla:1998rp}. This 3-dimensional CFT, as well as its dual
$AdS_4 \times S^7$ supergravity  solution, preserve the maximal number
of 32 supersymmetries, and exhibit $SO(2,3) \times SO(8)$ global symmetry.
While the M2-brane world-volume CFT is still not completely understood,
significant progress has been made over the past few years in constructing
a Lagrangian realization in which all or most of the symmetries and
supersymmetries are realized explicitly
\cite{Bagger:2006sk,Bagger:2007jr,Gustavsson:2007vu,Aharony:2008ug}.

\sm

The insertion of local and/or non-local gauge-invariant operators in the
CFT will break some or all of the supersymmetries and global symmetries,
and  lead to supergravity duals with correspondingly reduced symmetries.
An exciting theoretical laboratory is provided by supergravity solutions with
16 residual supersymmetries (or half-BPS solutions). A general classification
via semi-simple Lie superalgebras was given in \cite{D'Hoker:2008ix}
for  half-BPS solutions in M-theory which are locally asymptotic to
$AdS_4 \times S^7$. An analogous
classification for solutions locally asymptotic to $AdS_7 \times S^4$ in M-theory, and
locally asymptotic to $AdS_5 \times S^5$ in Type IIB were also obtained there.
These supergravity solutions form families in which the corresponding space-time
exhibits a warped $AdS$-factor, and they are dual to scale-invariant quantum field
theories. Remarkably, the corresponding half-BPS equations may often be
reduced to  integrable systems, which may then be solved completely, and exactly.
The solutions exhibit a wealth of topological and metrical structure, and are
characterized by interesting, and generally complicated,  moduli spaces.

\sm

It was shown in  \cite{D'Hoker:2008wc} that all solutions to M-theory which are
half-BPS, exhibit global $SO(2,2) \times SO(4)\times SO(4)$ symmetry, and
are locally asymptotic to either $AdS_4 \times S^7$ or to  $AdS_7 \times S^4$,
may be constructed via a warped product $AdS_{3} \times S^{3} \times S^{3}$
over a 2-dimensional Riemann surface $\Sigma$, and a certain integrable system on $\Sigma$.  Physically, these solutions will be relevant as AdS/CFT duals to CFTs with
various arrangements of defects and/or interfaces.\footnote{Half-BPS solutions
in M-theory with space-time manifold $AdS_{3} \times S^{3} \times S^{3} \times T^2$  
were constructed in \cite{Boonstra:1998yu, Gauntlett:1998kc,deBoer:1999rh}. 
Other types of solutions on various space-times with an $AdS_{3}$ factor, and 
various degrees of supersymmetry, have been constructed in 
\cite{Gauntlett:2006qw,Donos:2008ug}. For an earlier derivation of the BPS 
equations in M-theory for the warped geometry used in this paper, see 
\cite{Yamaguchi:2006te,Lunin:2007ab}.} The solutions of \cite{D'Hoker:2008wc}
are {\sl local supergravity solutions} in the sense that the associated bosonic 
supergravity fields satisfy the Bianchi and field equations locally, and the
corresponding BPS equations allow for 16 independent spinor solutions, locally.
The local solutions may or may not be globally regular or physical (e.g. 
the metric may fail to be real throughout).

\sm

To  obtain regular physically acceptable solutions (such as those allowed to enter
the AdS/CFT correspondence), one must impose global regularity
and boundary conditions. For the half-BPS solutions locally asymptotic
to $AdS_7 \times S^4$ (which were referred to as cases II and III in
\cite{D'Hoker:2008wc}), suitable global regularity and boundary conditions
were obtained in \cite{D'Hoker:2008qm}, and the resulting globally regular
solutions were constructed using a simple linear superposition principle.
Each solution is invariant under the superalgebra $OSp(4^*|2) \oplus OSp(4^*|2)$.
The resulting families of solutions are labelled by an integer $g \geq 0$, and
the solutions within each family are described by $2g + 1$ independent real moduli.
These solutions are fully back-reacted supergravity solutions dual  to  the
supersymmetric self-dual string solution of the 6-dimensional $(2, 0)$
super-symmetric M5-brane world-volume theory.

\sm

The purpose of the present paper is to impose suitable regularity and
boundary conditions on the local  half-BPS solutions of \cite{D'Hoker:2008wc}
which are locally asymptotic to $AdS_4 \times S^7$ (referred to as case I there), 
and to investigate their existence. Solutions of this type are invariant under the 
superalgebra $OSp(4|2,\bR) \oplus OSp(4|2,\bR)$.
One new class of such  solutions was obtained in \cite{D'Hoker:2009gg}, 
and referred to as the {\sl M-Janus solution}.
It consists of a 1-parameter family of deformations of $AdS_4 \times S^7$
which give a holographic realization of a Janus-like defect/interface CFT. 
The existence of Janus-like solutions in M-theory is surprising, 
as M-theory contains no dilaton field. Still, in \cite{D'Hoker:2009gg},
these solutions were obtained analytically, and were found to be regular.
Thus, the goal of this
paper reduces to investigating {\sl the existence of deformations of 
$AdS_4 \times S^7$ beyond those of the M-Janus solution.}

\sm

The regularity and boundary conditions for half-BPS solutions locally asymptotic
to either $AdS_4 \times S^7$ or $AdS_7 \times S^7$ are as follows.
Locally, all such solutions may be constructed in terms of a real, positive harmonic
function $h$ on a Riemann surface $\Sigma$, and a complex-valued function
$G$ on $\Sigma$ which satisfies a first order partial differential equation
\cite{D'Hoker:2008wc},
\bea
\label{G1}
\p_w G = \half (G + \bar G) \, \p_w \ln h
\eea
for an arbitrary local complex coordinate system $w, \bar w$ on $\Sigma$.
This partial differential equation, which is common to all cases I, II, and III,  is
to be supplemented with a non-linear algebraic constraint, and with
boundary conditions  which are case-dependent \cite{D'Hoker:2008wc}. These
conditions are as follows, 
\begin{description}
\item[Asymptotic to $AdS_4 \times S^7$] (case I)
\begin{itemize}
\item local regularity condition: $|G|^2 > 1$ inside $\Sigma$;
\item boundary conditions: $h=0$, and $G = + i$ or $G=-i$ on $\p \Sigma$.
\end{itemize}
\item[Asymptotic to $AdS_7 \times S^4$] (cases II and III)
\begin{itemize}
\item local regularity condition: $-4 |G|^4 - (G-\bar G)^2  > 0 $ inside $\Sigma$;
\item boundary conditions: $h=0$, and $G = 0$ or $G=i$ on $\p \Sigma$.
\end{itemize}
\end{description}
For fixed $h$, the partial differential equation (\ref{G1}) is {\sl linear} in $G$ (under
linear superpositions of $G$ with real coefficients). Locally on $\Sigma$,
a complex coordinate system may be chosen for which $h = \Im (w)$.  The
differential equation (\ref{G1}) is then invariant under translations of $\Re (w)$,
and may be solved in terms of Fourier transforms of combinations of modified
Bessel functions \cite{D'Hoker:2008wc}. Even locally, however, the algebraic constraints on $G$ still need to be point-wise enforced, making the problem
effectively non-linear. 

\sm

In the case of solutions which are locally asymptotic to $AdS_7 \times S^4$, 
nonetheless, the harmonic function $h$
could be used as a global coordinate. Physically, this was possible because 
the supergravity solutions have only a single $AdS_7 \times S^4$ asymptotic 
region. The function $G$ could be obtained by a simple linear superposition 
and, remarkably, the algebraic constraints turned out to be automatically obeyed  \cite{D'Hoker:2008qm}\footnote{By contrast, half-BPS solutions  Type IIB supergravity
\cite{Lin:2004nb,D'Hoker:2007xy,D'Hoker:2007xz,D'Hoker:2007fq},
may be systematically characterized by meromorphic (or harmonic) functions
and forms on $\Sigma$, fixed uniquely by boundary conditions.}.

\sm

In the case of solutions which are locally asymptotic to $AdS_4 \times S^7$, however, 
the boundary conditions listed above require the space-time to contain multiple 
asymptotic $AdS_4 \times S^7$ regions. In this case, $h$ is generally not 
a good global coordinate. Therefore,  the Riemann surface $\Sigma$, the 
harmonic function $h$, and the function $G$, {\sl must all be considered as 
unknowns}. This situation renders the system for solutions asymptotic to 
$AdS_4 \times S^7$ ``more non-linear" than the one for the $AdS_7 \times S^4$
 case, thus more complicated, with physically different results.

\sm

Surprisingly, we find that for locally asymptotic $AdS_4 \times S^7$
boundary conditions, the only regular deformations of the M-Janus
solutions are of a topological nature, and produce multiple covers
of the basic M-Janus solution. In particular, we shall show
analytically that no non-trivial continuous infinitesimal deformations exist away from
the M-Janus solution, and give numerical evidence that no
finite deformations exist either. These arguments are presented
within the context of a set of mild assumptions on the structure of
the general regular solutions, which we regard as natural.\footnote{In 
\cite{Lunin:2007ab}, the existence was advocated  of regular solutions 
which have the same symmetries as we have imposed, 
obey  $AdS_{4}\times S^{7}$ asymptotics, and support 4-form charge.
The existence of such solutions, if actually different from our multi-covered M-Janus 
solutions, would be in contradiction with our results. 
The origin of the discrepancy between \cite{Lunin:2007ab} and our results 
is not entirely clear at this time, though we note that the analysis of \cite{Lunin:2007ab} 
appears not to involve any quadratic constraint, whose importance was 
paramount in our work, and that a full local solution is not available in \cite{Lunin:2007ab}.}
However, the possibility of solutions containing a localized singularity 
corresponding to the insertion of a singular M5-brane into $AdS_4 \times S^7$ 
is still an open problem.

\sm

The absence, on the AdS side, of regular half-BPS deformations away 
from the M-Janus solution has a surprising consequence on the dual CFT side.
It implies the non-existence of half-BPS interface/defect operators,
in the dual $2+1$-dimensional maximally supersymmetric CFT, which 
would correspond to such deformations. In ABJM theory, 24 supersymmetries 
are manifest on the local fields of the QFT for all values of $k$, and one
would expect many interface/defect operators which preserve 12 
supersymmetries. The remaining 8 supersymmetries of the theory for $k=1,2$
are not manifest but emerge from non-local ``monopole operators", or more
accurately from instantons in $2+1$dimensions. Enhancing the 
12 supersymmetries of any interface/defect operator to 16 should probably
again be caused by instantons, but their geometry may or may 
not be consistent with the presence of an interface.  The results of the current 
paper  suggest that such enhancements to interface/defect operators
with 16 supersymmetries should not exist. We plan to investigate
these CFT questions in future work.

\sm

Throughout the course of our investigations into the existence of
solutions to the differential equation (\ref{G1}), we provide various reformulations
of the problem, which may be of interest in their own right. First, we show that
equation (\ref{G1}) is equivalent to an $SL(2,\bR)$-invariant Helmholtz-like,
or automorphic, equation. Second, we show that the general solution of the 
differential equation (\ref{G1}), and its related automorphic equation,  may be
reformulated in terms of the {\sl Hermitian pairing of a number of meromorphic 
functions (or meromorphic blocks) on $\Sigma$}. Third, we show that the general solution to
the differential and algebraic constraint equations for solutions asymptotic 
to $AdS_7 \times S^4$ (obtained in \cite{D'Hoker:2008qm}) are given by the 
{\sl Helgason transform} corresponding to a certain non-unitary representation of 
$SL(2,\bR)$. Finally, we show that the solutions to the differential equation
(\ref{G1}) for the case asymptotic to $AdS_4 \times S^7$ also correspond
to a non-unitary representation of $SL(2,\bR)$, but whose role in mathematics
appears to be, thus far, unclear. Similarly, the role in $SL(2,\bR)$-representation 
theory of the non-linear constraints remains to be elucidated.

\sm

The structure of the remainder of this paper is as follows. In section 2, we shall
review the local solution, the regularity and the boundary conditions for solutions
which are locally asymptotic to $AdS_{4}\times S^{7}$. In section 3, we shall
review the M-Janus solutions, and construct their multiple covers.
In section 4, we shall show that the general local solution may be expressed
in terms of {\sl Hermitian pairings} of  certain  meromorphic functions,
and discuss the $SL(2,\bR)$ group theoretic underpinnings of the
$AdS_4 \times S^7$ as well as of the $AdS_7 \times S^4$ solutions.
In section 5, we shall show that the general natural Hermitian pairing structure for
$AdS_4 \times S^7$ always leads to configurations that violate the
quadratic constraint $|G|^2 >1$.  In section 6, we shall interpret our
rigidity results in terms of the dynamics of M2- and M5-branes.
Finally, in section 7, we shall conclude with a discussion of loose ends,
open problems and questions for future investigations. Two technical 
discussions are relegated to Appendices A and B.

\newpage

\section{Local solution, regularity, and boundary conditions}
\setcounter{equation}{0}
\label{localsol}

In this section, we review the local half-BPS solutions obtained in
\cite{D'Hoker:2008wc}.
(Derivations may be found in \cite{D'Hoker:2008wc}; they  will not be needed here,
and will not be repeated.) The 11-dimensional metric Ansatz consists of a fibration
of $AdS_{3}\times S^{3}\times S^{3}$ over a 2-dimensional Riemann surface 
$\Sigma$ with boundary $\p \Sigma$,
\bea
\label{metricansatza}
ds^2 = f_1^2 ds_{AdS_3}^2 + f_2^2 ds_{S_2^3}^2 + f_3^2 ds_{S_3^3}^2 +  ds_{\Sigma}^2
\eea
The 4-form field strength is given by
\bea
\label{fluxansatz}
F_{4} = g_{1a}\;  \omega_{AdS_{3}} \wedge e^{a}+
g_{2a}\;  \omega_{S_2^{3}} \wedge e^{a}+
g_{3a}\; \omega_{S_3^{3}} \wedge e^{a}
\eea
where  $\omega_{AdS_{3}}$ and $\omega_{S^{3}_{2,3}}$ are the volume forms on
$AdS_{3}$ and $S^{3}_{2,3}$ respectively, and  $e^{a}, a=1,2$  is an orthonormal
frame on $\Sigma$.  In terms of an arbitrary system of local complex coordinates
$w,\bar w$ on $\Sigma$, the metric on $\Sigma$ in (\ref{metricansatza}) reduces
to the standard conformal form,
\bea
\label{rhofactor}
ds_{\Sigma}^2= 4 \rho^{2 } |dw|^2
\eea
The metric factors $f_1, f_2, f_3, \rho$, as well as the flux fields
$g_{1a}, g_{2a}$, and $g_{3a}$,  only depend on  $\Sigma$.
The Ansatz automatically respects $SO(2,2)\times SO(4)\times SO(4)$ symmetry,
which may also be viewed as the symmetry of an AdS/CFT dual 1+1-dimensional
conformal interface or defect in the 3-dimensional M2-brane CFT.

\sm

In \cite{D'Hoker:2008wc}, the BPS equations governing solutions with 16
residual supersymmetries were reduced to constructing a Riemann surface
$\Sigma$ with boundary, a real positive harmonic function $h$ on $\Sigma$,
and the solution to a first order partial differential equation on $\Sigma$ for
a complex-valued field $G$, subject to a point-wise non-linear algebraic constraint.
(The origin of $G$ in terms of Killing spinor components, and its relations to
these spinors may be found in \cite{D'Hoker:2008wc}.)
The partial differential equation for $G$ is given by,
\bea
\label{gequation}
2 \p_w G = (G + \bar G) \p_w \ln h
\eea
for an arbitrary complex coordinate system $w, \bw$ on $\Sigma$.
For solutions locally asymptotic to $AdS_4 \times S^7$ (case I), the field $G$
is subject to the following point-wise quadratic constraint,
\bea
\label{qc}
|G(w, \bw)|^2 > 1 \qquad {\rm for~all} \qquad (w,\bw) ~ \hbox{on the inside of} ~  \Sigma
\eea
For any given harmonic function $h$, the differential equation (\ref{gequation})
is linear in the sense that if $G_1$ and $G_2$ are two solutions of
(\ref{gequation}), then so is $\lambda _1 G_1 + \lambda _2 G_2$, where
$\lambda _1, \lambda_2$ are {\sl real} constants. Since the complex
coordinates $w, \bar w$ are arbitrary, we may choose locally
$\Im (w) = h$, and then solve the equation (\ref{gequation}) by Fourier
analysis, as was done in  \cite{D'Hoker:2008wc}. To solve the full set of reduced
BPS equations, however, it remains to enforce the point-wise quadratic
constraint (\ref{qc}), which poses a highly non-trivial problem.

\subsection{The real field $\Phi$}

In all generality, the equation (\ref{gequation}) for $G$ may be partially integrated
in terms of a single real function. To see this, we multiply (\ref{gequation}) on both
sides by the anti-holomorphic 1-form $\p_{\bar w} h$. This gives the following equation,
\bea
\p_w \left ( G \p_{\bar w} h \right ) =  (G+\bar G) {|\p_w h|^2 \over 2 h}
\eea
Since the right hand side of this equation is real, we have
$\p_w \left ( G \p_{\bar w} h \right ) = \p_{\bar w} \left ( \bar G \p_w h \right )$.
As a result there exists, at least locally, a real function $\Phi$ such that
$G (\p_{\bar w} h) = \p_{\bar w} \Phi$. Actually, the Riemann surfaces
of interest to us here will always be contractible, and $\Phi$ will be singularity-free
on the inside of $\Sigma$, so that the local result will hold globally on $\Sigma$.
Thus, $\Phi$ provides a partial integral of (\ref{gequation}), as well as
an economical parameterization of $G$, given by,
\bea
G = { \p_{\bar w} \Phi \over \p_{\bar w} h}
\eea
Equation (\ref{gequation}) expressed in terms of $\Phi$ takes the form,
\bea
\label{Phieq}
2 \p_w \p_{\bar w} \Phi - \p_{\bar w} \Phi (\p_w \ln h) - \p_w \Phi (\p_{\bar w} \ln h)=0
\eea
In terms of the field $\Phi$, the quadratic constraint (\ref{qc}) becomes
an inequality on the derivatives of $\Phi$, given by
$|\p_w \Phi |^2 \geq |\p_w h|^2$.

\subsection{Metric factors and anti-symmetric tensor in terms of $G$ and $h$}

In order to express in terms of $h$ and $G$ the local half-BPS solutions which 
are locally asymptotic to $AdS_4 \times S^7$, and investigate their regularity 
and boundary conditions,
it will be useful to define the following real function $W^2$ on $\Sigma$,
\bea
W^2 \equiv 4 |G|^4 + (G - \bar G)^2
\eea
Assuming that $|G|^2 \geq 1$, we automatically have $W^2 \geq 0$.
The metric factors in (\ref{metricansatza}) are then expressed as follows,
\bea
\label{metricfactors}
f_{1}^{6}
	&=&
	{ h^{2} W^{2} \over 16^2 (|G|^{2}-1)^{2}}
\no \\
f_{2}^{6}
	&=&
	 h^{2}{(|G|^{2}-1)\over 4 \, W^{4}} \left ( 2 |G|^{2} + i (G-\bar G) \right )^{3}
\no \\
f_{3}^{6}
	&=&
	 h^{2}{(|G|^{2}-1)\over 4 \, W^{4}} \left ( 2 |G|^{2} - i (G-\bar G) \right )^{3}
\eea
The metric factor in (\ref{rhofactor}) is given by,
\bea
\rho^{6}
&=&
{\left | \partial_{w} h\right | ^6 \over 16^2 h^{4}} \big(|G|^{2} -1 \big) W^{2}
\eea
The anti-symmetric tensor field-strengths are expressed in terms of $g_i$. 
They can be  defined by conserved currents as follows,
\bea
\label{bi}
(f_1)^3 g_{1w} = - \p_w b_1
&=&  { 3 W^2 \p_w h \over 32 G (|G|^2 -1)}
- { 1+G^2 \over 16 G (|G|^2-1)^2} \, J_{w}
\no\\
(f_2)^3 g_{2w} = - \p_w b_2 
&=& 
-{ (G+i)  ( 2 |G|^2  + i (G-\bar G)  )^2 \over W^4} \, J_w
\no\\
(f_3)^3 g_{3w} = - \p_w b_3 
&=& 
+ { (G-i)  ( 2  |G|^2  -i (G -\bar G)   )^2 \over W^4} \, J_w
\eea
where the following quantity was used for notational compactness,
\bea
J_{w} = \half  (G \bar G - 3 \bar G^2 + 4 G \bar G^3) \p_w h + h G  \p_w \bar G
\eea
 It was shown in \cite{D'Hoker:2008wc} that the equations of motion of  as well as the Bianchi identities are satisfied for a harmonic $h$ and a $G$ which solves (\ref{gequation}).

\subsection{The $AdS_4 \times S^7$ solution}

The simplest solution is the maximally symmetric $AdS_4\times S^7$ itself.
The Riemann surface is the infinite strip,
\bea
\Sigma=\{ w \in \bC, ~ w = x+iy, ~  x \in \bR, \; 0\leq y\leq \pi/2\}
\eea
Note that the Riemann surface $\Sigma$ has two boundary components.
In these coordinates, the functions  $h$ and $G$ for the $AdS_4 \times S^7$ solution
are given by,
\bea
\label{hgads}
G = i {\ch (w + \bar w) \over \ch (2 \bar w)}
& \hskip 1in &
\left \{ \matrix{h =   8 \, \Im (  \sh (2w)  ) \cr
\Phi  =  - 8 \sh (w + \bar w)  \cr } \right .
\eea
Using  (\ref{metricfactors}), the metric factors become,
\bea
\label{adssol}
f_1 = \ch(2 x)
\qquad
f_2 = -2 \cos (y)
\qquad
f_3 = -2 \sin(y)
\qquad \qquad
\rho = 1
\eea
The boundary is characterized by the vanishing of the harmonic function $h=0$,
or alternatively, by $G = \pm i $.
On the lower boundary of the strip, where $y=0$, one has $G=+i$, which implies
that the radius $f_2$  of $S_2^3$ vanishes. On the upper boundary of the strip,
where $y=\pi/2$, one has  $G=-i$, which implies that the radius $f_3$ of $S_3^3$
vanishes.  The boundary of $AdS_4 \times S^7$ on the other
hand is located at $x = \pm \infty$.
An additional piece of the $AdS_4 \times S^7$ boundary is located at the $1+1$-dimensional boundary of the $AdS_3$ fiber, along which the $x = + \infty$ and $x = -\infty$ pieces of the boundary are glued together.

\subsection{General Regularity and boundary conditions}

Physically interesting solutions generally require solutions to be everywhere
regular and to be locally asymptotic to $AdS_4 \times S^7$. This leads to
the following three assumptions for the geometry of each solution.

\begin{enumerate}
\itemsep -0.01in
\item
The boundary of the 11-dimensional geometry is locally asymptotic to
$AdS_4 \times S^7$.
\item
The metric factors are finite everywhere on $\Sigma$, except at points where the
geometry becomes locally asymptotic to $AdS_4 \times S^7$, in which case
the $AdS_3$ metric factor diverges.
\item
The metric factors are everywhere non-vanishing, except on the boundary
$ \p \Sigma$, in which case at least one sphere metric factors vanishes.
In addition, the $S_2^3$ and $S_3^3$ metric factors may vanish simultaneously
only at isolated points on $\p \Sigma$.
\end{enumerate}
The second requirement guarantees that all singularities in the geometry are locally
of the same type as $AdS_4 \times S^7$.  The third requirement guarantees
that the boundary of $\Sigma$ corresponds to an interior line in the
11-dimensional geometry.

\subsection{Analysis of the regularity and boundary conditions}

It follows from  (\ref{metricfactors}) that a particular combination of metric factors is
very simple
\bea
\big(f_{1} f_{2} f_{3}\big)^2= h^2
\eea
The metric factor $f_{1}$ is given by a positive definite expression in terms
of spinor coordinates,  and cannot vanish  \cite{D'Hoker:2008wc}.
Hence the  condition $h=0$ (which defines a 1-dimensional subspace in
$\Sigma$) occurs if and only if at least one of the metric factors for the
spheres $f_{2}$ or $f_{3}$ vanishes. It follows from  assumption $3.$
that $h=0$ defines the boundary of $\Sigma$.

\sm

Next,  we analyze the boundary conditions which $G$ has to satisfy at $h =0$.
On the domain for $G$, namely $|G|^2 \geq 1$, we automatically have
$W^2\geq 0$. Furthermore, if $|G|^2 >1$ then we have $W^2 >0$. Vice-versa,
 if $W^2=0$, then we have $G = \pm i$.
In order to study the boundary conditions, it is useful to
exhibit the following combinations,
\bea
{f_2^2 + f_3^2 \over f_1^2 } & = & { 16 |G|^2 (|G|^2-1) \over W^2}
\no \\
{f_2 ^2 \over f_2^2 + f_3^2} & = & \half  + { i \over 4 |G|^2} (G-\bar G)
\no \\
{f_3 ^2 \over f_2^2 + f_3^2} & = & \half  - { i \over 4 |G|^2} (G-\bar G)
\eea
$\bullet$ In the region $|G|\geq 1$, each normalized sphere metric factor vanishes
at exactly one point,
\bea
{f_2^2  \over f_2^2 + f_3^2} = 0 & \qquad \Leftrightarrow \qquad & G= +i
 \no \\
{f_3 ^2 \over f_2^2 + f_3^2} = 0 & \Leftrightarrow & G= -i
\eea
$\bullet$ As $h \to 0$, the metric factors $f_1^2$ and $f_2^2 + f_3^2$
never vanish,  so that we must have
\bea
h \to 0  \hskip 0.5in \left \{ \matrix{ |G|^2 -1 & \sim & h ^2 \cr
 W & \sim & h \cr} \right .
\eea
$\bullet$ As $ h \to \infty$, the metric factors $f_2$ and $f_3$ remain finite,
while $f_1$ blows up, so that,
\bea
h \to \infty  \hskip 0.5in \left \{ \matrix{
W & \sim & {\rm finite} \cr
|G|^2-1 & \sim & W^4/ h^2 \cr} \right .
\eea
In summary, on the boundary $\p \Sigma$, we have $h=0$, and $G$ alternates
between the values $\pm i$. In the interior of $\Sigma$, we must have $h >0$
and $|G|>1$.

\newpage

\section{M-Janus solution and its multiple covers}
\label{secthree}
\setcounter{equation}{0}

It was shown in \cite{D'Hoker:2009gg} that the maximally symmetric
$AdS_4 \times S^7$  solution has a simple deformation to a half-BPS
solution of the Janus type, or simply the {\sl M-Janus solution}.
The surface $\Sigma$ is given by the same strip as the $AdS_4 \times S^7$
solution was, namely
$\Sigma=\{ w \in \bC, ~ w = x+iy, ~  x \in \bR, \; 0\leq y\leq \pi/2\}$,
while the functions $h$, $\Phi$ and $G$ now depend on an arbitrary
real parameter $\lambda$, and a positive constant $h_0$, and are given by,
\bea
h & = & - 4i  h_0   \Big ( \sh (2w) - \sh (2 \bar w) \Big )
\no \\
\Phi & = & - 8 h_0
\Big ( \sh (w + \bar w) - \lambda \ch (w+\bar w) \Big )
\no \\
G & = & i { \ch (w + \bar w) + \lambda \sh (w- \bar w ) \over \ch (2 \bar w)}
\eea
The maximally symmetric  $AdS_4 \times S^4$ solution corresponds to
$\lambda =0$. The radius $R_0$ of the asymptotic $AdS_4$ space of
the solution is given by
\bea
R_0 ^6 = { h_0 ^2 \over 1 + \lambda ^2}
\eea
The resulting metric factors and flux fields may be
found in \cite{D'Hoker:2009gg}, and will not be needed here. The M-Janus 
solution is everywhere regular, and has two distinct asymptotic
$AdS_4 \times S^7$ regions. The 2+1-dimensional CFT dual results from
the maximally supersymmetric CFT through the insertion of a 1+1-dimensional
linear interface/defect, which partially breaks the full $OSp(8|4,\bR)$
superconformal symmetry.

\subsection{The M-Janus solution in upper half-plane coordinates}

The upper half-plane provides a more uniform system of coordinates
for $\Sigma$ that will lend itself better to generalizing the M-Janus
solutions. Using the change of variables $u =  \th (w)$, the $w$-strip
is mapped onto the $u$-upper half-plane $\Sigma = \{ u \in \bC, \, \Im (u) >0 \}$,
in terms of which the functions $h$, $\Phi$, and $G$ become,\footnote{In the
sequel, we shall set  $h_0 =1/16$, for convenience.}
\bea
\label{Mjanus}
h & = & 8 i h_0 \left ( { 1 \over u+1} + {1 \over u-1} - {\rm c.c.} \right )
\no \\
\Phi & = & 16 h_0 { u + \bar u - \lambda ( u \bar u -1) \over |u^2-1|}
\no \\
G & = & i \left ( { \bar u^2 -1 \over |u^2-1|} \right )
{ u \bar u +1 + \lambda (u - \bar u) \over \bar u ^2+1}
\eea
The quadratic constraint is automatically obeyed for all values
of $\lambda$, since we have,
\bea
|G|^2 - 1 = 4 (1 + \lambda ^2) {   \Im (u)^2  \over |u^2+1|^2}
\eea
which is strictly positive for $\Im (u) >0$. At $u=\infty$, the functions
$h$, $\Phi$, and $G$ are regular.
The two distinct asymptotic  $AdS_4 \times S^7$ regions of the
M-Janus are located at $u=\pm 1$, namely the poles of $h$.
Traversing the real $u$ axis, from $-\infty$ to $+\infty$, the value of
the function $G$ starts as $+i$, changes to $-i$ upon crossing $u=-1$,
and resumes the value $+i$ upon crossing $u=+1$. Thus, we have
$f_2=0$ for $|u| > 1$, and $f_3 =0$ for $|u|<1$.

\subsection{The Ansatz with multiple $AdS_4 \times S^7$ regions}

In seeking to construct solutions with multiple (more than 2)
asymptotic $AdS_4 \times S^7$ regions, we are naturally led to
associating the different poles of $h$ with the different asymptotic
regions. The surface $\Sigma$ will be taken to be the upper half-plane
parametrized by a {\sl global} complex coordinate system $u, \bar u$.
The boundary $\p \Sigma$ is then the real axis. The harmonic
function $h$ must vanish on $\p \Sigma$. As a result, the poles $a_i$ of $h$
must be on the real axis, and the residues at these poles must be real.
It will be convenient to leave the point at infinity as a regular point of the
solution manifold. Thus, we have the following general form for $h$,
\bea
\label{h}
h
= { i \over 2} \sum _{i=1}^{2g+2} \left ( { c_i \over u - a_i}  - { c_i \over \bu - a_i} \right )
=  \sum _{i=1}^{2g+2}  { c_i \, \Im (u) \over |u - a_i|^2}
\eea
The harmonic function $h$ now automatically vanishes for $u$ real.
Regularity of the supergravity solutions inside $\Sigma$ requires that $h >0$
on the inside of the upper half-plane. From the second equality in (\ref{h}),
it is manifest that this condition is equivalent to,
\bea
c_i > 0 \hskip 1in i = 1, \cdots , 2g+2
\eea
Vanishing residues reduce the Ansatz to one with lower $g$.
Equivalently, the harmonic function may be recast in the following form,
\bea
\label{harmdef}
h  = { i \over 2} \left (
{R(u) \over Q(u) } - {R(\bu) \over Q(\bu)} \right )
\eea
where $Q$ is a real polynomial of degree $2g+2$ and $R$ is a real
polynomial of degree $2g+1$. Positivity of the residues now takes the
form, $c_i=R(a_i)/Q'(a_i) \geq 0$. The polynomials $Q$ and $R$ may also
be immediately  derived from (\ref{h}), and are given by
\bea
\label{QR}
Q(u) = \prod _{i=1} ^{2g+2} (u-a_i)
\hskip 1in
R(u) = \sum _{i=1}^{2g+2} c_i \prod _{j\not= i} (u-a_j)
\eea
For the M-Janus solution in (\ref{Mjanus}), we have $g=0$, and the field $\Phi$
takes the form of a degree 2 polynomial in $u$ and $\bu$, divided by the
absolute value of the function $Q(u)=u^2-1$. For general $g$, we shall
postulate that $\Phi$ is of the form,
\bea
\label{phans}
 \Phi (u, \bu) = {P(u, \bar u) \over |Q(u)|}
\eea
where $Q(u)$ is given by (\ref{QR}), and $P$ is a real polynomial, i.e.
satisfying $\overline{P(u, \bu)} = P(u, \bu)$. Inspection of the M-Janus solution
shows that the total degree in $u$, $\bar u$ of $P$ must be $2g+2$,
a result which is further confirmed by the fact that  $u =\infty$
is then a regular point of $\Phi$, as it is in the M-Janus solution.
The Ansatz (\ref{phans}) also guarantees that near any one of the poles
$a_i$, the field $\Phi$ has a singularity of the form, $\Phi \sim 1 / |u-a_i|$,
producing an $AdS_4 \times S^7$ asymptotic region near each pole.

\subsection{Reduced differential equation and boundary conditions \label{secphirest}}

With the above Ansatz of (\ref{phans}),  the differential equation of (\ref{Phieq})
for $\Phi$ is equivalent to a polynomial relation between $P, R$, and $Q$,
given by,
\bea
\label{polyrel}
0 & = &
4 (Q \bar R - R \bar Q) \p_u \p_{\bar u} P
+ \Big (  ( \p_u Q) (\p_{\bar u} \bar R) - (\p_u R) ( \p_{\bar u} \bar Q)  \Big ) P
\no \\ &&
+2  \Big ( (\p_u R) \bar Q -  (\p_u Q) \bar R \Big ) \p_{\bar u} P
+2 \Big (R (\p_{\bar u} \bar Q) - Q (\p_{\bar u} \bar R)  \Big ) \p_u P
\eea
where we have used the abbreviations, $Q = Q(u)$,
$\bar Q = Q(\bar u)$, $R=R(u)$, and $\bar R= R(\bar u)$.
In terms of these functions, $G$ takes the following form,
\bea
\label{G1chap3}
G = i { \bar Q \over |Q|} \left ( { P \p_{\bar u} \bar Q - 2 \bar Q \p_{\bar u} P
\over \bar Q \p_{\bar u} \bar R - \bar R \p_{\bar u} \bar Q} \right )
\eea
If $P, R, Q$ satisfy the polynomial relation (\ref{polyrel}) then $G$, given above,
will automatically  satisfy the partial differential equation (\ref{gequation}).

\sm

The explicit expression for $G$ in terms of $P,Q$, and $R$ of (\ref{G1chap3})
allows us to recast the boundary conditions for $G$ in simple terms.
The boundary conditions for $G$ are that $G$ can take only the values
$+i$ or $-i$ on $\p \Sigma = \bR$, with the sign alternating precisely when
the boundary coordinate $u \in \bR$ crosses one of the  poles $a_i$ of $h$.
Restricting $u$ to the real axis in $G$ amounts to setting $\bar u =u$.
By construction, the phase pre-factor $i \bar Q/|Q|$ of $G$ in (\ref{G1chap3}) alternates
between $+ i$ and $-i$ as the coordinate $u$ crosses the points $a_i$ on the real axis.
Thus, the boundary condition on $G$ is equivalent to the requirement that the
remaining (rational) factor of $G$ in (\ref{G1chap3}) be equal to $-1$ throughout the
real axis. This condition translates to the following condition for all $u=\bu \in \bR$,
\bea
P (u, \bu ) \p_\bu Q( \bu )  - 2  Q(u)  \p_\bu P(u, \bu )
=  R(\bu) \p_{\bar u}  Q(\bu) -Q(\bu) \p_{\bar u}  R(\bu)
\eea
Evaluating this expression at the zeros $u=\bar u = a_i$ of $Q(u)$,
for $i=1, \cdots, 2g+2$, simplifies the expression, and gives (after omitting a
common non-vanishing factor $Q'(a_i)$),
\bea
\label{bc1}
P(a_i, a_i)  =  R(a_i) = c_i Q'(a_i)
\eea
Since $P(u,u)$ and $R(u)$ are polynomials in $u$ of degree $2g+2$ and $2g+1$
respectively, which agree at $2g+2$ points, these polynomials must differ
by a polynomial of degree $2g+2$ which vanishes on all $2g+2$ points $a_i$,
and which must thus be a multiple of $Q(u)$. We conclude that we must have
the following relation,
\bea
\label{bc2}
P(u,u) = R(u) + d_\infty Q(u)
\eea
The real constant $d_\infty$ is determined by matching the
$u \to \infty$ limits of $P(u,u)$ and $Q(u)$. It is straightforward to show
that the M-Janus solution is the most general solution to these equations
and boundary conditions for the special case $g=0$.

\subsection{Solving the equations for $g=1$}

We now analyze a general solution with $g=1$, for which the harmonic
function $h$  has four poles on the real axis, suitable for solutions with four
distinct asymptotic $AdS_4 \times S^7$ regions.

\sm

By an $SL(2,R)$ transformation, we place three of the four poles  at $u=-1,0,+1$,
respectively. The position of the fourth pole is then an arbitrary point on the
real axis, which will be denoted by $a$ with $a \not= 0, \pm 1$.  The harmonic
function $h$ is parameterized in terms of the following functions, 
defined in (\ref{QR}),
 \bea
Q(u)&=& (u+1) u (u-1) (u-a)
\no\\
R(u)&=& +c_{1} u(u-1)(u-a) +c_{2} (u+1)(u-1)(u-a)
\no\\
&& +c_{3}u (u+1)(u-a) +c_{4}(u+1)u (u-1)
\eea
Here, $c_{i}>0$ with $i=1,2,3,4$ are the residues of the harmonic function $h$
at its poles $a_i$. We use the results of section \ref{secphirest}, and parametrize $P$ as
follows,\footnote{The form of $P$ presented here is not the most general real polynomial of degree 4; the proof that all other terms may be omitted is given in Appendix A.}
\bea
P(u,\bar u)
&=&
2d_{0,0}+ d_{1,0}(u+\bar u) + d_{2,0} (u^{2}+\bar u^{2}) +2  d_{1,1} u\bar u
\no\\
&&
+ d_{2,1}(u^{2}\bar u + \bar u^{2}u) +2d_{2,2}u^{2}\bar u^{2}
\eea
The residues $c_i$ are obtained in terms
of the polynomial $P$ via relations (\ref{bc1}), and we have,
\bea
\label{crelfour}
c_{1}&=&
(d_{0,0}- d_{1,0}+d_{1,1}+d_{2,0}-d_{2,1}+d_{2,2})/( a+1)
\no\\
c_{2}&=&
-2 d_{0,0}/a
\no\\
c_{3}&=&
(d_{0,0}+ d_{1,0}+d_{1,1}+d_{2,0}+d_{2,1}+d_{2,2}) / ( a-1)
\no\\
c_{4}&=&
- {2\big( d_{0,0}+a  d_{1,0}+a^{2 }d_{1,1}+a^{2}d_{2,0}+a^{3}d_{2,1}+a^{4}d_{2,2}\big)
\over a(a-1)(a+1)}
\eea
The terms of order $(u^m \bar u^n + u^n \bu^m)$ with $0\leq m,n\leq 4$
in (\ref{polyrel}) produce 25 equations, which are linear in $c_{i}$ and $d_{m,n}$.
Eliminating the $c_{i}$ from these equations employing (\ref{crelfour})
produces 25 quadratic equations. Remarkably only three equations
are linearly independent, all others are either trivially satisfied or linear
combinations of the following equations,
\bea
0&=&
4 a \,d_{0,0} d_{1,1}-a \,d_{1,0}^{2}+2 d_{1,0}d_{2,0}
+ 2 a\, d_{1,1}d_{2,0}+ 2a\, d_{2,0}^{2}- 4 d_{0,0}d_{2,1}
\no\\
&&- a \,d_{1,0}d_{2,1}- 8a\, d_{0,0}d_{2,2}\;\;
\label{deqone}
\\
0&=&
2 d_{1,0}^{2}- 8 d_{0,0} d_{1,1}+ a\, d_{1,0} d_{2,0}
-2a\, d_{0,0}d_{2,1}+ a\, d_{2,0}d_{2,1}-2 a\, d_{1,0}d_{2,2}
\label{deqtwo}
\\
0&=&
4 d_{1,0}d_{2,0}+ 2a d_{1,1}d_{2,0}+ 2a\,d_{2,0}^{2}
- 8 d_{0,0}d_{2,1}-a\, d_{1,0}d_{2,1}- 2 d_{2,0}d_{2,1}-a\,d_{2,1}^{2}
\no\\
&& -8a\,d_{0,0}d_{2,2}+ 4 d_{1,0}d_{2,2}+ 4a\, d_{1,1}d_{2,2}
\label{deqthree}
\eea
One can solve equations (\ref{deqone}) and (\ref{deqtwo}) for $d_{2,2}$ and
$d_{2,1}$ since they are linear in these unknowns. Substituting this result into
equation (\ref{deqthree}) yields a single remaining equation, which may be
factorized into 5 factors, four of which are linear in the variables $d_{m,n}$,
and one of which is bilinear,
\bea
\label{deqfour}
0&=&
\Big( d_{1,0}+ a (2d_{0,0}+d_{2,0})\Big)\Big( 2(2+a)d_{0,0}+ a(d_{1,0}-d_{2,0})\Big)
\\
&&\times \Big( 2(2-a) d_{0,0}+ a(d_{1,0}+d_{2,0})\Big)\Big(d_{1,0}+ a(-2 d_{0,0}+d_{1,1}+d_{2,0})\Big) \Big(d_{1,0}^{2}- 4 d_{0}d_{1,1}\Big)
\no
\eea
Each one of the five factors corresponds to a different branch of solutions.

\sm

It is easy to show that the fifth branch, where the fifth factor in (\ref{deqfour})
vanishes, i.e. $d_{1,0}^{2}- 4 d_{0}d_{1,1}=0$ leads to a solution where
$|G|=1$ everywhere. This solution produces a singular metric and is unphysical.

\subsection{All $g=1$ regular solutions are double covers of M-Janus}

The first branch is given by the vanishing of the first factor in (\ref{deqfour}),
$d_{1,0}+ a (2d_{0,0}+d_{2,0})=0$.
This equation, together with (\ref{deqone}) and (\ref{deqtwo}) may be used to
express  $d_{2,2},d_{2,1}$ in terms of $d_{0,0},d_{1,0},d_{1,1}$, as well as
the position  $a$ of the fourth pole,  and we find,
\bea
d_{2,2}&=& d_{0,0}+ d_{1,0}/a +d_{1,1}/ a^{2}
\no\\
d_{2,1}&=& -d_{1,0}-2 d_{1,1}/a
\no\\
d_{2,0}&=& -2 d_{0,0}- d_{1,0}/a
\eea
Remarkably, all these relations are linear in the coefficients $d_{m,n}$.
Inserting these relation into (\ref{crelfour}), one obtains the following
expressions for the residues $c_{i}$,
\bea
\label{cresult}
c_{1}&=& (1+a)d_{1,1}/ a^{2}
\no\\
c_{2}&=& -2 d_{0,0}/a
\no\\
c_{3}&=&  (a-1) d_{1,1}/ a^{2}
\no\\
c_{4}&=& -2(a^{2}-1) d_{0,0} / a
 \eea
Regularity requires $h >0$ inside $\Sigma$, which imposes the condition $c_{i}>0$,
for all $i=1,2,3,4$. Since we have $c_1c_3 = (a^2-1)d_{1,1}^2 /a^4$, and
$c_2 c_4 = 4 (a^2-1) d_{0,0}^2 /a^2$, the requirement of $c_i >0$ imposes
the condition $|a|>1$. It remains then only to require that $c_1>0$ and $c_2>0$,
or simply $a d_{0,0} <0$, and $0 < a d_{1,1} $. It may be readily checked that
the quadratic constraint $|G|^2  \geq 1$ is indeed satisfied for any choice of
$c_1>0$ and $c_2 >0$. Near each one of the poles of $h$, namely $-1,0,+1,a$,
the solutions have $AdS_4 \times S^7$ asymptotic behavior.

\sm

Remarkably, the general $g=1$ solution obtained above is
actually a double cover of the M-Janus solution. To see this, we introduce
the following two functions,
\bea
\phi_1 (u) = \left ( { u(u-a) \over u^2-1} \right )^\half
\hskip 1in
\phi_2 (u) = \left ( { u^2-1 \over u(u-a)} \right )^\half
\eea
which are, of course, inverses of one another. The general $g=1$
solution may be expressed in terms of $\phi_1, \phi_2$, as follows,
\bea
h   & = & -i {d_{1,1} \over a^2} \phi_1^2 - i d_{0,0} \phi_2^2
+ {\rm c.c.}
\no \\
\Phi  & = & 2{d_{1,1} \over a^2} \left | \phi _1  \right |^2
+ 2 d_{0,0} \left | \phi _2  \right |^2 + { d_{1,0} \over a} \left (
\phi_1^* \phi _2 + \phi _2 ^* \phi _1 \right )
\eea
The M-Janus solution of (\ref{Mjanus}) may be slightly generalized
by observing that its two poles at $\pm 1$ remain invariant under
a 1-parameter subgroup of $SL(2,\bR)$. Upon applying this group
of transformations to (\ref{Mjanus}), we find its most general presentation,
in terms of a general coordinate $v$ of the upper half-plane,
\bea
h & = &
{i \over 4} \left ( (\delta _{0,0} + \delta _{1,1} - \delta _{1,0} ){ v-1 \over v+1}
+ (\delta _{0,0} + \delta _{1,1} + \delta _{1,0}) { v+1 \over v-1} -{\rm c.c.} \right )
\no \\
\Phi & = &
{ 2 \delta _{0,0} + \delta _{1,0} (v+\bar v) + 2 \delta _{1,1} v \bar v \over |v^2-1|}
\eea
where the residues satisfy $\delta _{0,0} + \delta _{1,1} - \delta _{1,0} <0$, and
$\delta _{0,0} + \delta _{1,1} + \delta _{1,0}>0$.
The M-Janus solution is now seen to be equivalent to the general $g=1$
solution, upon setting,
\bea
\phi _1 (v) = \left ( { v-1 \over v+1} \right )^\half
\hskip 1in
\phi _2 (v) = \left ( { v+1 \over v-1} \right )^\half
\eea
and matching the free parameters as follows,
\bea
\delta _{0,0} & = & d_{0,0} + d_{1,1} / a^2 - d_{1,0} /a
\no \\
\delta _{1,0} & = & 2 d_{0,0} - 2 d _{1,0} / a
\no \\
\delta _{1,1} & = & d_{0,0} + d_{1,1} / a^2 + d_{1,0} / a
\eea
This matching of the $g=1$ and the M-Janus (or $g=0$) solutions was made
at the cost, however, of performing a change of variables $u \to v(u)$,
\bea
{v-1 \over v+1} = {u(u-a) \over u^2-1}
\eea
which is conformal throughout the upper half-plane, except at isolated
points where the derivative $\p v / \p u = 2 (au^2-2u+a)/(au-1)^2$ vanishes
or diverges. For $u$ in the upper half-plane, these points are given by,
\bea
au = 1 \hskip 1in au =   1 + i \sqrt{a^2-1}
\eea
The map $u \to v(u)$ is 2 to 1, in the sense that every point $v$ in the
upper half-plane is the image of two distinct points $u$ in the upper half-plane.
It is in this sense that the general $g=1$ solution is the double cover of
the M-Janus solution.

\sm

Above, we have examined only the first branch of the $g=1$ solutions,
corresponding to the vanishing of the first factor in (\ref{deqfour}).
The other remaining 3 branches correspond to other regions for $a$,
and may be solved in an analogous manner. For each branch, the corresponding
$g=1$ solution again maps onto a double cover of the M-Janus solution.

\subsection{General $g$ solutions and multiple covers of M-Janus}

For $g \geq 2$, the number of unknown coefficients in $P(u,\bu)$
grows rapidly, and neither MAPLE nor MATHEMATICA appear to
be capable of solving the corresponding equations in general.
For every $g \geq 2$ there exists, however, a family of solutions
analogous to the ones we found for $g=1$, amounting to a $g+1$-fold
cover of the M-Janus solution. The structure of the solution is as follows,
\bea
\label{solg}
Q(u) & = & Q_1(u) Q_2(u)
\no \\
R(u) & = & d_1 Q_1(u)^2 + d_2 Q_2(u)^2  - (d_1+d_2) Q_1(u) Q_2(u)
\no \\
P(u,\bu) & = &
d_1 \left | Q_1 (u) \right |^2 + d_2 \left | Q_2 (u) \right |^2
+ d_0 \left ( Q_1(u) Q_2(\bu) + Q_1 (\bu) Q_2(u) \right )
\eea
Here, $d_0, d_1, d_2$ are real constant coefficients. The function
$Q(u)$ is given by the customary expression in terms of the poles
$a_i$ of the harmonic function $h$.
The functions $Q_1(u)$ and $Q_2(u)$ provide a partition of the
$2g+2$ zeros $a_i$ of $Q(u)$ into two groups of $g+1$ zeros, in an
alternating order. To produce an explicit formula, it is convenient to
prescribe a definite order to the zeros, which we shall choose to be,
\bea
\label{order}
a_1 < a_2 < a_3 < \cdots < a_{2g} < a_{2g+1} < a_{2g+2}
\eea
The functions $Q_1(u)$, $Q_2(u)$ are built  from the
zeros with odd and even indices respectively,
\bea
Q_1 (u) = \prod _{i=1} ^{g+1} \left ( u - a_{2i-1} \right )
\hskip 1in
Q_2 (u) = \prod _{i=1} ^{g+1} \left ( u - a_{2i} \right )
\eea
The residues $c_i$ of the poles in the harmonic function $h$,
are given as follows,
\bea
i=\hbox{odd} & \hskip 1in & c_i = d_2 { Q_2(a_i) \over Q_1'(a_i)}
\no \\
i=\hbox{even} & \hskip 1in & c_i = d_1 { Q_1(a_i) \over Q_2'(a_i)}
\eea
Thanks to the alternating order of the zeros of $Q_1(u)$ and $Q_2(u)$,
the condition of positivity on $c_i$, for all $i=1, \cdots, 2g+2$,
simply reduces to the requirements,
\bea
\label{dsign}
0 < d_1 \hskip 1in d_2 <0
\eea
The non-linear constraint $|G|^2 >1$, on the inside of $\Sigma$,
is then automatically satisfied for all values of $d_0$. One could check
that this functional form satisfies the differential equation for $P$,
by direct calculation. It is more instructive, however, to show that
this solution is a multiple cover of M-Janus, which will then automatically
guarantee that it also satisfies the differential equation.

\sm

The correspondence with M-Janus may be exhibited explicitly,
by defining the functions,
\bea
\phi_1 (u) = \left ( { Q_1(u) \over Q_2(u) } \right ) ^\half
\hskip 1in
\phi_2 (u) = \left ( { Q_2(u) \over Q_1(u) } \right ) ^\half
\eea
Clearly, we have $\phi _1 (u) \phi _2(u) =1$. The functions
$h$ and $\Phi$ for the $g\geq 2$ solution may be recast as follows,
\bea
h & = & - {i \over 2} \left ( d_1 \phi _1^2 + d_2 \phi _2^2 - {\rm c.c.} \right )
\no \\
\Phi & = & 2d_1 |\phi _1|^2 + 2d_2 |\phi_2|^2
+ d_0 ( \phi_1 ^* \phi _2 + \phi _2 ^* \phi_1 )
\eea
It is now straightforward to read off the correspondence with the M-Janus
solution, just as we did for the case $g=1$, and we find here,
\bea
{ v -1 \over v + 1 } = {Q_1(u) \over Q_2(u)}
\eea
as well as the following correspondence of the parameters,
\bea
\delta _{0,0} & = & d_1 + d_2 - d_0
\no \\
\delta _{1,0} & = & 2d_2 - 2 d_0
\no \\
\delta _{1,1} & = & d_1 + d_2 + d_0
\eea
The map $u \to v(u)$ now fails to be conformal at multiple points,
given by the poles and zeros of the derivative,
\bea
{\p v \over \p u} = 2 { Q_1Q_2' - Q_1' Q_2 \over (Q_1 - Q_2)^2}
\eea
Since $Q_1$ and $Q_2$ are both monic polynomials of degree $g+1$,
and have alternating zeros, the derivative $\p v/\p u$ has $g$ (double)
poles on the real axis, and $2g$ complex zeroes, of which $g$ are in the
upper half-plane. The map $ u \to v(u)$ is $g+1$ to 1, and thus
produces a $g+1$-fold cover of the M-Janus solution.

\newpage

\section{Hermitian pairing}
\setcounter{equation}{0}

In the preceding section, we generalized the family of M-Janus solutions,
using an Ansatz of rational functions with $g+2$ asymptotic $AdS_4 \times S^7$
regions. The resulting solutions, however, invariably reduced to multiple
covers of the M-Janus solution. In particular, no new solutions arose
for which we have non-trivial fluxes and charges on the $S^3$-spheres.
In the present section, we shall generalize the solution, and show that
any solution of the differential equation for $G$ is given by a {\sl Hermitian
pairing form}, namely,
\bea
\label{hermitian}
h (u, \bu) & = & { i \over 2} \Big ( \kappa (u) - \overline{\kappa (u)} \Big )
\no \\
\Phi (u, \bu) & = & \sum _{m,n=1} ^N D_{mn} \overline{\phi_m (u) } \phi _n (u)
\eea
where $\kappa$ and $\phi_m(u)$ are meromorphic functions of $u$, and
$D^t =D$ is a real symmetric matrix, assumed to be of maximal rank $N$.
The number $N$ of ``meromorphic blocks" can be finite or infinite.
Using the arguments of section 3.3, the meromorphic function $\kappa (u)$ 
may be expressed in terms of the remaining data as follows,
\bea
\label{kappa}
\kappa (u) = \sum _{m,n=1} ^N D_{mn} \phi _m(u) \phi _n (u)
\eea
The $AdS_4 \times S^7$ solution, its M-Janus generalization, and
the multiple covers found in the preceding section, all fit into the
Hermitian pairing form with $N=2$. Conversely, one can show
that these solutions constitute the most general solution to the
problem with $N=2$.

\subsection{Reduction to complex analytic equations}

The power of the Hermitian pairing reformulation
is that it allows us to reduce the differential equation for $\Phi$,
and the boundary conditions for the associated $G$, to purely
complex analytic equations. The corresponding derivation is
given in Appendix B, and we limit ourselves here to quoting the
final result. We use matrix notation, in which $\phi_m$ with
$m=1,\cdots, N$ are grouped into a column matrix, and $D$
is the square matrix with entries $D_{mn}$.

\sm

{\sl The Hermitian pairing form of (\ref{hermitian}) satisfies the
differential equation for $\Phi$, and the boundary conditions for $G$,
provided the following relation holds,
\bea
\label{Heq}
(\p _u \kappa ) D \phi - 2\kappa D \p_u \phi = 2H \p_u \phi
\eea
for some real symmetric  matrix $H$, which is independent of $u$.
(Proven in Appendix B.)}

\sm

Assuming that $H$ is known, and using the fact that $D$ and $H$ are
$u$-independent matrices, it is straightforward to obtain the general
local solution for $\phi$ of (\ref{Heq}), and we find,
\bea
\label{sol1}
\phi (u) = \left ( \kappa + D^{-1} H \right ) ^\half \phi _0
\eea
where $\phi_0$ is a column matrix of integration constants. The easiest
way to derive this result is by choosing local
complex coordinates in which $u = \kappa$, and using the fact that
$D$ is of maximal rank and thus invertible, to obtain the equation
$2 (\kappa +   D^{-1} H ) \p_\kappa \phi = \phi$, which is readily integrated.
The relation between $\kappa$ and $\phi _m$ of (\ref{kappa})
produces relations between $D$, $H$, and $\phi _0$, given by,
$\kappa = \phi ^t D \phi = \phi _0 ^t \left ( \kappa D + H \right ) \phi _0 $, or
\bea
\label{sol2}
\phi _0 ^t D \phi _0 & = & 1
\no \\
\phi _0 ^t H \phi _0 & = & 0
\eea
The above solution is local in the sense that $\kappa$ provides a good
local coordinate system, but may not extend to a good global coordinate.
The presence of the square root in the solution will introduce further
global coordinate issues, which will have to be dealt with. To gain a better
understanding of the problems involved, we begin by diagonalizing $D^{-1}H$.

\subsection{Diagonalizing $D^{-1}H$}

Since $D$ and $H$ are real symmetric matrices, the product $D^{-1}H$
will be real, but not, in general, symmetric. We shall assume
here that the eigenvalues of $D^{-1}H$ are all mutually distinct, so that
$D^{-1}H$  may be properly diagonalized (i.e. not requiring the Jordan
normal form). The diagonalization matrix $S$ will, in general, not be
an orthogonal matrix. We set,
\bea
D^{-1} H = S X S^{-1} \hskip 1in \det (S) \not= 0
\eea
where $S$ is independent of $u$, just as $D$ and $H$ are, and
$X$ may be thought of as a diagonal matrix, though its precise form
will be specified later. Upon defining,
\bea
\phi (\k ) & = &  S \psi (\k)
\no \\
\phi _0 & = &  S \psi _0
\eea
the equations of (\ref{sol1}), and (\ref{sol2})  become,
\bea
\label{fund2}
\psi (\kappa ) = \left ( \kappa + X \right )^\half \psi _0
\hskip 1in
\left \{ \matrix{ \psi ^t _0 \left ( S^t H S \right ) \psi _0 & = & 0 \cr
\psi ^t _0 \left ( S ^t D S \right ) \psi _0 & = & 1 \cr}
\right .
\eea
With $X$ diagonal, the system of square roots is now completely
decoupled from one another, and may be examined eigenvalue by
eigenvalue.

\sm

There is a subtlety that we shall now address and resolve.
If the eigenvalues of $D^{-1}H$ are all real, then $X$ can be taken
to be properly diagonal, and the matrix $S$ can be chosen to be real.
If some of the eigenvalues are complex, however, they will occur in
complex conjugate pairs (since the matrix $D^{-1}H$ is real), and the
matrix $S$ will no longer be real. This situation would be problematic,
because throughout we have assumed the basic functions $\psi _m$
and $\phi_m$ to be real functions. When this is the case, $X$ can be  
block-diagonalized,
by separating the real eigenvalues $x_1, \cdots, x_p$ from the
complex eigenvalues $w_1 = u_1 + i v_1 , \cdots, w_q = u_q + i v_q$,
and their complex conjugates $\bar w_1, \cdots, \bar w_q$,
\bea
X = \left ( \matrix{ X_r & 0 \cr 0 & X_c \cr} \right )
\eea
where
\bea
X_r & = & \hbox{diagonal}(x_1,  \cdots, x_p)
\no \\
X_c & = & \hbox{diagonal}(u_1+v_1\ep,  \cdots, u_q+v_q\ep)
\hskip 0.6in \ep = \left ( \matrix{0 & 1 \cr -1 & 0 \cr} \right )
\eea
and $p+2q=N$. The matrix $X$ is now  real,
even though some of its eigenvalues appear in complex conjugate pairs.
In this basis, the matrix $S$ can be chosen to be real as well.

\subsection{Hermitian pairing holds for the general solution}

In the above discussion, the Hermitian pairing form (\ref{hermitian}) of the solution
was introduced as an Ansatz which naturally generalizes the form
of the M-Janus solution and its multiple covers. In this subsection, we shall
show that the Hermitian pairing form in fact holds for the general
solutions for $h$, $\Phi$, and $G$ to the differential equations.

\sm

Our starting point is the general form of the solution to the differential
equations for $G$, derived already in \cite{D'Hoker:2008wc}. We shall cast
it in the following form,
\bea
\kappa = \tilde h - i h
& \hskip 1in & 
\Phi = \Phi _1 + \Phi _2
\no \\ &&
G = G_1 + G_2
\eea
where $h$ and $\tilde h$ are harmonic, $G_i = \p_{\bu} \Phi_i/(\p_\bu h)$
for $i=1,2$, and $G_1, G_2$ are given by
\bea
G_1 & = & h \int _{-1} ^1 { dt \over \sqrt{1-t^2} } \left [
(1-t) C_1'(th+i \tilde h) + (1+t) \overline{C_1'(th + i \tilde h) } \right ]
\no \\
G_2 & = & h \int _0 ^\infty { dt \over \sqrt{t^2-1} } \left [
(1-t) C_2'(th+i \tilde h) + (1+t) \overline{C_2'(th + i \tilde h) } \right ]
\eea
Here, $C_1'$ and $C_2'$ are meromorphic functions
of their arguments. For $C_1'$, the argument runs throughout the
complex plane (since $t$ can be either positive or negative, and $\tilde h$
runs through the real line). For $C_2'$, the argument runs through a
half-plane (since $t$ is now positive only). Note that $C_1'$ and $C_2'$
are NOT holomorphic functions of either $\kappa$ or of $\bar \kappa$.
As $h \to \infty$, the function $G$ must remain finite for all values of $\tilde h$.
As a result, the functions $C_1'$ and $C_2'$ must tend to 0 as their
argument tends to $\infty$.

\sm

It was shown in \cite{D'Hoker:2008wc}
that both the $AdS_4 \times S^4$ and $AdS_7 \times S^4$ solutions
corresponds to combinations of simple poles of the functions $C_1'$
and $C_2'$. Thus, we begin by investigating the
solutions $G$ and $\Phi$ produced by poles.

\subsubsection{$G$ and $\Phi$ produced by poles in $C_1'$}

A pole in $C_1'$ is of the form,
\bea
C_1 '(v) = { a \over v +b}
\eea
There being no a priori reality restrictions on $C_1'$, the parameters
$a$ and $b$ will generally take complex values.
Using the following formula, already derived in \cite{D'Hoker:2008wc},
\bea
\label{lambda2}
{ a h (1-t) \over th + i \tilde h +b} + { \bar a h (1+t) \over th - i \tilde h + \bar b} =
-a + \bar a + a { z+1 \over z+t} - \bar a { \bar z -1 \over \bar z +t}
\eea
where
\bea
\label{lambda1}
z= { i \tilde h + b \over h}
= { \k + \bar \k - 2 i b \over \k - \bar \k}
\eea
and the following integral formula,
\bea
\int _{-1} ^1 { dt \over \sqrt{1-t^2}} \, { 1 \over t +z} = { \pi \over \sqrt{z^2-1}}
\eea
we find,
\bea
\label{G1}
G_1 =  a \pi \left ( { \k -ib \over \bar \k -ib} \right )^\half
- \bar  a \pi \left ( { \k + i\bar b \over \bar \k +i \bar b} \right )^\half
\eea
Here, we have dropped an irrelevant imaginary additive constant
$-\pi (a - \bar a) $  to $G_1$. For $\k $ real, the function $G_1$ is
piece-wise constant, and thus suitable for obeying the boundary
conditions $G= \pm i$ (or $G=0,\pm i$ for the $AdS_7 \times S^4$
deformations).
From $G_1$, we derive the function $\Phi_1$, and we find,
\bea
\Phi _1= -i a \pi (\k - ib)^\half (\bar \k - ib)^\half
+ i \bar a \pi (\k + i \bar b)^\half (\bar \k + i\bar b)^\half
\eea
By construction, the function $\Phi_1$ is real.

\subsubsection{$G$ and $\Phi$ produced by poles in $C_2'$}

The contribution of a simple pole in $C_2'$, given by,
\bea
C_2 ' (u) = { a \over v +b}
\eea
may be handled similarly, using equations (\ref{lambda2}) and (\ref{lambda1}),
but with the difference that the relevant  integral formula to be applied now is,
\bea
\int _1 ^\infty { dt \over \sqrt{t^2-1 }} \, { 1 \over t +z}
= { 1 \over \sqrt{z^2-1}} \ln \Big ( z+ \sqrt{z^2-1} \Big )
\eea
Omitting an irrelevant (divergent) purely imaginary constant contribution, we
find,
\bea
G_2 = a \left ( { z+1 \over z-1} \right )^\half
\ln \Big ( z+ \sqrt{z^2-1} \Big )
- \bar a \left ( { \bar z-1 \over \bar z + 1} \right )^\half
\ln \Big ( \bar z+ \sqrt{\bar z^2-1} \Big )
\eea
In contrast with the function $G_1$ produced by a pole in $C_1'$, the
function $G_2$ is NOT piecewise constant as $\kappa$ runs over the real axis.
In fact, the real axis must be approached with care as the complex
function $z$, defined in (\ref{lambda1}), will diverge as $\kappa - \bar \kappa \to 0$.
We approach the real axis from above, and choose the following
parameterization to do so, $\k = i \ep + x$ for $x$ real and $\ep >0$.
As $\ep \to 0$, it is clear from (\ref{lambda1}) that $z \to \infty$,
\bea
z =  -(ix +b)/  \ep
\eea
As $\ep \to 0$, the square root factors tend to $\pm 1$, and are
piecewise constant as a function of $x$, but the logarithmic factors
do depend on $x$, resulting in the following limit,
\bea
\lim _{\ep \to 0} G_2
=  \pm a \ln \Big ( -2ix - 2b \Big ) \mp \bar a \ln \Big ( 2ix - 2 \bar b \Big )
\eea
which depends on $x$.  Thus, the function $G_2$ arising from a single
pole in $C_2'$ cannot satisfy the piecewise constant boundary
conditions which the supergravity problem requires for $G$.
The only option is to cancel this $x$-dependence on the
boundary by adding the contribution with the following interchange
$(a,b) \to (\bar a, - \bar b)$, giving a pole structure of the form,
\bea
C_2 '(v) = { a \over v+b} + { \bar a \over v - \bar b}
\eea
It is straightforward to assemble all the pieces for this combination
of poles in $C_2'$. Using the branch cut discontinuity relation
$\ln \zeta - \ln (-\zeta) = - i \pi$, we find,
\bea
G_2 = - i \pi a  \left ( { \k -ib \over \bar \k -ib} \right )^\half
-i \pi  \bar  a \left ( { \k + i \bar b \over \bar \k +i \bar b} \right )^\half
\eea
Remarkably, this result coincides with the result obtained for a single
pole in $C_1'$, namely the expression given in (\ref{G1}), after a
simple redefinition to $ a \to i a$.

\subsection{Mapping pole solutions to Hermitian pairing}

In summary, a simple pole in $C_1'$ and a pair of simple poles in $C_2'$
both result in the same $G$-functions which are piecewise constant
on the real axis, and are thus compatible with the boundary conditions
required for $G$. Meromorphic functions $C_1'(v)$ or $C_2'(v)$ in the
plane $v \in \bC$, which tend to 0 as $|v| \to \infty$ must be sums of poles.
The resulting functions $G$ and $\Phi$ are linear combinations of these
basic solutions, and take on the  following form,
\bea
G & = & i \sum _{m=1}^p A_m  \left ( { \k -x_m \over \bar \k -x _m} \right )^\half
+ i \sum _{n=1}^q \left ( B_n  \left ( { \k -w_n \over \bar \k -w_n} \right )^\half
+ \bar  B_n \left ( { \k - \bar w_n \over \bar \k - \bar w_n } \right )^\half \right )
\\
\Phi & = & \sum _{m=1} ^p A_m |\kappa - x_m|
+ \sum _{n=1}^q \left ( B_n (\k - w_n)^\half (\bar \k - w_n)^\half
+  \bar B_n (\k - \bar w_n)^\half (\bar \k - \bar w_n)^\half \right )
\no
\eea
where $A_m$ and $x_m$ are arbitrary real parameters, related to the
data of the real poles, and $B_n$ and $w_n$ are arbitrary complex parameters
(with $\Im (w_n) >0$,  related to the data of the complex poles.

\sm

The calculations of the preceding two subsections demonstrate that the 
contribution of each pole to $\Phi$ is given by the  combination,
\bea
\label{Phiw}
\Phi _w (u, \bu) = ( \k  - w  )^\half  ( \bar \k  - w  )^\half
\eea
This basic building block for the solutions to the differential equation for
$\Phi$ is precisely of the form of the Hermitian pairing structure
advocated in the preceding section. More precisely, the number $N=p+2q$ 
corresponds to the rank of the matrix $D$ in the Hermitian pairing formulas.
The points $w$ correspond to the eigenvalues of the matrix $D^{-1}H$, and the
functions $(\kappa - w)^\half$ correspond to the functions
$\psi (\kappa)$ which diagonalize $D^{-1}H$. Thus, the Hermitian pairing
Ansatz can actually be derived from the general solution of the
differential equations for $\Phi$ and $G$, and the data in the pairing
Ansatz has a precise correspondence with the positions of the
poles of the functions $C_1'$ and $C_2'$ of the general local solution.

\sm

Two cases may be distinguished, according to whether $w$ in
$\Phi _w (u, \bu)$ is real or complex. If $w$ is real, the point $w$
is on the boundary of $\Sigma$, while when $w$ is complex
(with $\Im (w) >0$) the point is in the upper half-plane, and thus
inside $\Sigma$. These cases were distinguished also at the
level of the data of the Hermitian pairing formulation, namely
corresponding to real or pairs of complex eigenvalues of $D^{-1}H$.

\subsection{Harmonic analysis on the upper half-plane}

The differential equations for $G$ and for $\Phi$ are covariant under
$SL(2, \bR)$ transformations. A convenient way to establish this is via
a rescaled field $\hat \Phi $, in terms of which the differential equation
(\ref{Phieq}) becomes,
\bea
\label{sigma}
\p_u \p_{\bar u} \hat \Phi - {3 |\p_u h|^2 \over 4 h^2} \hat \Phi =0
\hskip 1in
\hat \Phi \equiv {\Phi \over \sqrt{h}}
\eea
The action of $SL(2,\bR)$ may be exhibited explicitly in terms of
the meromorphic function $\kappa (u)$, which is related to $h$ by
$h = i (\k - \bar \k)/2$,
\bea
\k  \to \k ' = {\alpha \, \k  + \beta \over \gamma \, \k  + \delta }
\hskip 1in  \a \delta - \b \g =1
\eea
where $\a, \b, \g, \delta \in \bR$ are constants. Under these transformations,
$\hat \Phi$ transforms as a scalar, while the original  $\Phi$-field transforms
as the absolute value of a spinor. Choosing the local complex coordinate
$u = \kappa$, the equation (\ref{sigma}) becomes,
\bea
\label{sigma1}
(\k - \bar \k)^2 \p_\k \p_{\bar \k} \hat \Phi = - {3 \over 4} \hat \Phi
\eea
The operator $(\k - \bar \k)^2 \p_\k \p_{\bar \k}$ is the familiar
Laplacian on the upper half $\kappa$-plane. With the above sign
conventions, the Laplacian is a positive operator on square 
integrable functions on the upper half plane, corresponding to the
various unitary series of $SL(2,\bR)$.

\sm

Equation (\ref{sigma1}) reveals, however, that $\hat \Phi$ is an 
eigenfunction of the Laplacian with {\sl negative} eigenvalue $-3/4$. 
Thus, $\hat \Phi$ solves a problem of harmonic analysis on the upper 
half-plane, but for non-unitary representations of $SL(2,\bR)$. The 
elementary solution $\Phi _w$, introduced in (\ref{Phiw}), produces
the combination $\hat \Phi _w = \Phi _w /\sqrt{h}$, which
satisfies (\ref{sigma1}) for any value of $w$. Under $SL(2,\bR)$,
the $\hat \Phi _w$ transform under the representation
defined by
\bea
\label{SL2}
\hat \Phi _w (\k, \bar \k ) \to \hat \Phi _{w'} (\k ',\bar \k ')
= {\hat \Phi _w (\k ,\bar \k ) \over \g w + \delta}
\hskip 1in
w'= {\alpha \, w  + \beta \over \gamma \, w  + \delta }
\eea
For solutions asymptotic to $AdS_7 \times S^4$, the point $w$ is real,
giving a real transformation prefactor in (\ref{SL2}), while for solutions
asymptotic to $AdS_4 \times S^7$, the point $w$ is complex (not real)
and the prefactor is genuinely complex.

\subsection{Reconsidering the deformations of $AdS_7 \times S^4$}

The half-BPS deformations of the $AdS_7 \times S^4$ solution,
obtained in  \cite{D'Hoker:2008qm} have a natural interpretation
in the set-up of the preceding subsections.  They correspond to
purely real poles, for which $q=0$ and thus $p=N$,
and for which each term in $G$,
\bea
G = i \sum _{m=1}^p A_m \left ( { \k - x_m \over \bar \k - x_m} \right ) ^ \half
\eea
is a pure phase on the real $\k$ boundary $\p \Sigma$. This
solution also has a natural interpretation in terms of harmonic
analysis on the upper half-plane. This is most easily seen
in terms of the function $\hat \Phi = \Phi / \sqrt{h}$, which was
introduced in the preceding subsection. Its explicit expression
may be cast in the following form,
\bea
\hat \Phi = \sum _{m=1} ^q A_m S(\k, x_m) ^{-\half}
\hskip 1in
S(\k, x_m) = { i(\k -\bar \k)/2  \over |\k - x_m|^2}
\eea
The object $S(\k, x_m)$ is the Poisson kernel for the upper half-plane
(or the bulk-boundary propagator in AdS/CFT language).
A Theorem of Helgason \cite{Hel} (see also \cite{Terras}) asserts that 
all the eigenfunctions of the Laplacian on the upper half-plane,
\bea
(\k - \bar \k)^2 \p_\k \p_{\bar \k} \Psi _s = s(1-s) \Psi _s
\eea
are of the following form,
\bea
\Psi _s (\k, \bar \k)= \int _\bR dx \, f(x) \, S(\k, x)^s
\eea
where $f(x)$ is a real function, or distribution, for all complex
values of $s$. For the case at hand, we have
$s=-1/2$, so that $s(1-s) = -3/4$, and the $f(x)$ is a sum of Dirac
$\delta$-functions. These $\delta$-functions are required by the
boundary condition on the associated $G$-function. The Helgason theorem shows
that, given these boundary conditions on $G$, the corresponding
solutions for $\hat \Phi$, and thus for $\Phi$ and $G$ are unique.

\sm

As a result, we conclude that our earlier construction in \cite{D'Hoker:2008qm} 
of solutions  which are locally asymptotic to $AdS_7 \times S^4$, does indeed 
produce the most general possible solutions given the prescribed regularity and 
boundary conditions. It remains an open question, however, whether
the non-linear constraint $W^2>0$ inside $\Sigma$ also has a natural
$SL(2,\bR)$ group theoretic interpretation.

\subsection{Hermitian pairing form of the M-Janus solution}

The M-Janus solution corresponds to the case $N=2$, and more
specifically $p=0$ and $q=1$. Thus, it is characterized by a single
complex variable $w_1$, and a complex constant $B_1$. We may use
$SL(2,\bR)$ symmetry to map the point $w_1$ to be anywhere in the
upper half-plane.  To make contact with the earlier presentation of the
solution, we shall choose $w_1=i$, yielding the following form,
\bea
\Phi = B_1 (1+i \k )^\half (1+ i \bar \k )^\half +
B_1^* (1- i \k )^\half (1- i \bar \k )^\half
\eea
This expression is not well-defined on the upper half-plane, since
it involves a quadratic branch cut starting at the point $\k =i$.
Single-valuedness is restored by uniformizing $\kappa$ in terms
of a complex variable $u$,
\bea
\k = { 2 u \over u^2-1}
\eea
for which we have
\bea
\Phi = B_1 { |u|^2 + i (u +\bu) -1 \over |u^2-1|}
+ B_1^* { |u|^2 - i (u +\bu) -1 \over |u^2-1|}
\eea
which reproduces the M-Janus solution exactly with the identifications,
\bea
B_1 = -8h_0 (\lambda +i)
\eea
We see that the building
block of the complex eigenvalue solution are indeed the basic M-Janus
solutions (but not just the basic $AdS_4 \times S^7$ solution).

\sm

General solutions to the differential equations for $G$ or for $\hat \Phi$, 
with boundary conditions on $G$ suitable for $AdS_4 \times S^7$ asymptotics,
may be obtained by linear superpositions of the basic M-Janus fields.
From the point of view of $SL(2,\bR)$ representation theory, such 
linear superpositions are allowed. The key questions, for the present 
supergravity system, is whether the quadratic constraint $|G|>1$
can be satisfied for such linear superpositions. This problem will
be tackled in the remaining sections on this paper.

\newpage

\section{Rigidity of M-Janus and its multiple covers}
\setcounter{equation}{0}

The M-Janus solution has two distinct asymptotic $AdS_4 \times S^7$
regions, and results from a single pair of complex conjugate eigenvalues
of $D^{-1}H$ or, equivalently, from a single pole in $C_1'(v)$. We shall now
seek solutions with four distinct asymptotic $AdS_4 \times S^7$ regions,
and we study, to this end, solutions to the differential equation for $\Phi$
based on two pairs of complex conjugate eigenvalues of $D^{-1}H$ or,
equivalently, two poles $w_1, w_2$ in $C_1'(v)$. Using $SL(2,\bR)$, we
choose one of the complex poles $w_1$  to be at $ i$. The point $i$ is left
invariant under the  one-parameter subgroup,
\bea
\k \to \k ' = {\k \cos \theta - \sin \theta \over \k \sin \theta
+ \cos \theta}
\eea
for any real value of $\theta$. Using this residual symmetry, we choose
the second pole $w_2$ to be at $i /k$, where $k$ is real, and $0<k<1$.
The relevant relations are most easily expressed in terms of $\xi$,
defined by $w_2 = \tg (\xi)$; one then has $2\theta = \xi + \xi^*$, and $k^2 |\tg (\xi - \theta)|^{-2}$.

\sm

Adopting the above choices for the branch points, the general solution with
four asymptotic $AdS_4 \times S^7$ regions is specified by,
\bea
\Phi & = & B_1 (1+i \k )^\half (1 + i \bar \k )^\half
+ B_1^* (1-i \k )^\half (1 - i \bar \k )^\half
\no \\ &&
+ B_2 (1+i k \k )^\half (1 + i k \bar \k )^\half
+ B_2^* (1-i k \k )^\half (1 - i k \bar \k )^\half
\eea
The function $\Phi$ has branch cuts in the upper half-plane,
starting at $\kappa =i$, and $\k = i/k$. The corresponding $G$-function is given by
$G= 2 i \p_{\bar \k} \Phi$, and takes the following form,
\bea
G =  - B_1 \left ( {1+ i \k \over 1 + i \bar \k} \right )^\half
+ B_1^* \left ( {1- i \k \over 1 - i \bar \k} \right )^\half
- B_2 \left ( {1+ i k \k \over 1 + i k \bar \k} \right )^\half
+ B_2^* \left ( {1- i k \k \over 1 - i k \bar \k} \right )^\half
\eea
Reliable study of the the branch cuts and the different sheets of the
corresponding Riemann surface is achieved by uniformizing the
coordinate $\kappa$.

\subsection{Uniformization in terms of Jacobi elliptic functions}

The absolute values $|1\pm i \k |$ and $|1\pm ik  \k |$ are always well-defined
and single-valued. Thus, it remains to properly uniformize the square roots
$\sqrt{1+\k ^2}$ and $\sqrt{1+k^2\k ^2}$ by means of a single
well-defined coordinate. This is done with the help of Jacobi elliptic 
functions\footnote{Our notations and conventions will follow those of 
\cite{bateman, magnus}.}
\bea
 i \k & = & \sn (z ,k)
\no \\
\sqrt{1 + \k^2} & = & \cn (z ,k)
\no \\
\sqrt{1 + k^2 \k ^2} & = & \dn (z ,k)
\eea
where the Jacobi elliptic functions $\sn (z ,k)$, $\cn (z ,k)$, $\dn (z ,k)$ are
meromorphic functions of a single well-defined complex coordinate $z $.
Using the above uniformization, we find,
\bea
h & = &  \half \Big ( \sn (z,k) + \sn (\bz, k) \Big )
\no \\
\Phi & = &  B_1 \, \cn (z,k) \, { |1- \sn (z,k)| \over 1- \sn (z,k)}
+ B_2 \, \dn (z,k) \, {  |1- k \, \sn (z,k)| \over 1- k\, \sn (z,k)}
+ {\rm c.c.}
\eea
The resulting function $G$ is given by,
\bea
G  & =  &
-  B_1 \, { |1+ \sn ( z,k) | \over \cn (\bz,k) }
+  B_1^*\, { |1-  \sn (z,k) | \over \cn (\bz,k) }
\no \\ &&
-  B_2 \, {  |1 + k \, \sn (z,k) | \over \dn (\bz,k) }
+  B_2^* \, {  |1 - k\,  \sn (z,k) | \over \dn (\bz,k) }
\eea
Next, we determine the fundamental domain for $\Sigma$.
It is chosen so that $h> 0$ inside $\Sigma$, and $h=0$ on the
boundary $\p \Sigma$. The condition for the vanishing of
$h$ is given by $\sn (z, k) + \sn (\bz, k) =0$.
Using the addition theorem for the function $\sn$, applied to
the real and imaginary parts $x,y$ of $z=x+iy$, we find that,
\bea
\label{hell}
h = {  \sn (x,k) \, \dn ( y, k') \over
\cn (y, k')^2 +  k^2 \sn (x, k)^2\, \sn (y, k')^2}
\eea
where $k^2 + (k')^2=1$. Here, we have used the following standard relations,
\bea
\label{sncn}
\sn (iy,k) = i {\sn (y,k')\over \cn (y,k')}
\hskip 0.5in
\cn (iy,k) = {1\over \cn (y,k')}
\hskip 0.5in
\dn (iy,k) = {\dn (y,k') \over \cn (y,k')}
\eea
To solve for the fundamental domain, we inspect the
periodicity properties, the zeros and the poles of the elliptic functions,
listed in Table 1.

\begin{table}[htdp]
\begin{center}
\begin{tabular}{|c||c|c|c|} \hline
  & $\sn (z,k)$  & $\cn (z,k) $ & $\dn (z,k)$
\\ \hline \hline
$z+2K$  & $-\sn (z,k)$ & $-\cn (z,k)$ & $+\dn (z,k)$
\\ \hline
$z+ 2iK'$  & $+\sn (z,k)$ & $-\cn (z,k)$ & $-\dn (z,k)$
\\ \hline
zeros & $2n K + 2 i m K'$ & $(2n+1) K + 2 i m K'$ & $(2n+1)K + (2m+1)iK'$
\\ \hline
poles & $ 2nK + (2m+1)i K'$ & $ 2nK + (2m+1)i K'$ & $ 2nK + (2m+1)i K'$
\\ \hline \hline
\end{tabular}
\end{center}
\caption{Properties of Jacobi elliptic functions}
\label{table1}
\end{table}

The function $\dn (y,k')$ has no zeros on the real line, so that
the factor $\dn (y,k')\not=0$. Furthermore, since $\dn (0,k')=1$,
we see that we have $\dn (y, k')\geq 0$. The zeros of $\sn (x,k)$
are $x=2Kn$, for $n \in \bZ$, so that we also have $h=0$ whenever
$x=\Re (z ) = 2 K n$. Thus, the fundamental domain with $h > 0$
may be chosen as follows,
\bea
\Sigma = \Big \{ z=x+iy \in \bC, ~ \hbox{with} ~  0 \leq x \leq 2K;  ~  |y| \leq 2 K' \Big \}
\eea
In view of  the periodicity of all three Jacobi elliptic functions under shifts by the
full period $4 i K'$, the edges $y = \pm 2 K'$ are to be identified, and
$\Sigma$ has the topology of an annulus, or finite cylinder (see Fig 1).

\vskip .5in

\begin{figure}\label{fig1}
\centering
\includegraphics[ scale=0.85]{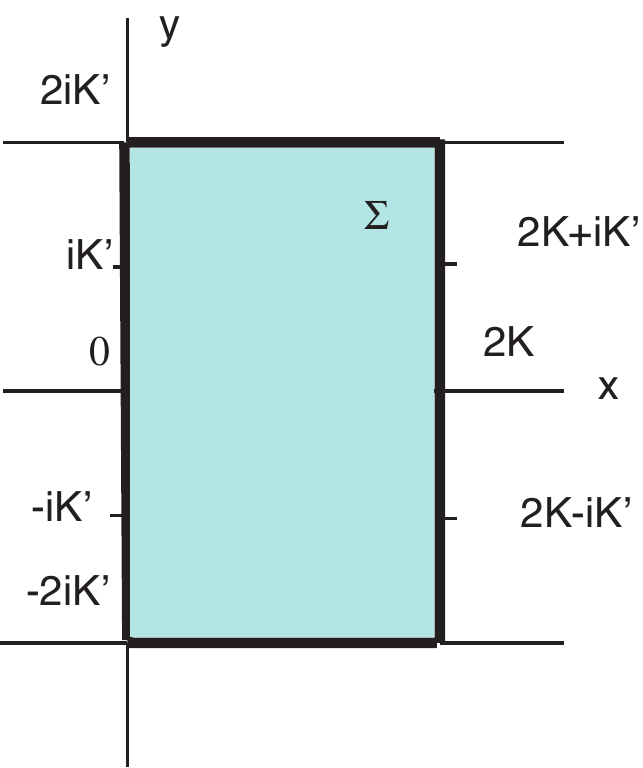}
\caption{The fundamental domain for $\Sigma$.}
\end{figure}

%

\subsection{Asymptotic regions and boundary conditions}

The poles of $h$ arise when the denominator of (\ref{hell}) vanishes,
which requires $\cn (y,k')=0$ and $\sn (x,k)=0$. Thus, there are no poles
on the inside of $\Sigma$. On the boundary $\p \Sigma$, there are
four distinct poles, namely at the points,
\bea
z_1 =  +i K' & \hskip 1in &  z_3 = 2K + i K'
\no \\
z_2 =  -i K' & \hskip 1in &  z_4 = 2K - i K'
\eea
To investigate the boundary conditions, we evaluate $G$ on the two
disjoint boundary components, parametrized by $z = i y$ and $z=2K+ i y$
both for $|y|\leq 2K'$. The expressions for $G$ are, respectively,
\bea
G(z=iy) & = & {\cn (y,k') \over |\cn (y,k') |} \Big (
-  (B_1 - B_1^*)  -  (B_2-B_2^*)    \Big )
\no \\
G(z=2K+iy) & = &  {\cn (y,k') \over |\cn (y,k') |} \Big (
+  (B_1 - B_1^*)   -  (B_2-B_2^*)   \Big )
\eea
Here, we have applied a number of simplifications, made with the help of relations
(\ref{sncn}), as well as using the fact that $\dn (y,k')>0$ for all real $y$.
Clearly, the real parts of $B_1$ and $B_2$ are unconstrained by the
boundary conditions, and the overall sign of the boundary conditions
may be reversed by reversing the imaginary parts of $B_1$ and $B_2$.
Thus, without loss of generality, we may choose $G(z=0)=+i$, which will
imply that $G=+i$ for $z=iy$ and $|y|<K'$ and $G=-i$ for $z=iy$
and $K'<|y|<2K'$. With this choice, we now have two possibilities left,
\bea
G(z=2K) = +i & \hskip 1in & B_1^* = B_1 \qquad \qquad B_2^*-B_2=+i
\no \\
G(z=2K) = -i & \hskip 1in & B_2^* = B_2 \qquad \qquad B_1^*-B_1=+i
\eea
Note that the entire boundary behavior is generated by either the $B_1$
terms, or the $B_2$ terms, but not both at the same time.

\subsection{The constraint $|G|^2 >1$}

Systematic numerical analysis shows that with the above boundary
conditions, the quadratic constraint $|G|^2>1$ is obeyed for all
$z \in \Sigma$, if and only if either $B_1=0$, or $B_2=0$.
In particular, whenever $B_1$ and $B_2$ are both non-zero,
there is always a region of $\Sigma$ where $|G|<1$.
In fact, we find an even stronger result, namely that
whenever $B_1$ and $B_2$ are non-zero, there exists
at least one point $z_0 \in \Sigma$ where $G(z_0)=0$.

\sm

It is possible to prove the above result analytically in the special
case of  small deformations away from the M-Janus solution.
These are achieved, for example,  by keeping $B_1$ fixed,
and taking $B_2$ to be small (compared to $B_1$). The $B_2$-terms
then introduce only small changes in $G$ over most of the
fundamental domain, except in the neighborhood of the
poles of the $B_2$-terms. The function $G$ has the following poles in
$\Sigma$,
\bea
\hbox{$B_1$-terms} & \hskip 1in & z = K , \, K + 2iK'  \equiv K - 2iK'
\no \\
\hbox{$B_2$-terms} & \hskip 1in & z = K +  iK', \, K - iK'
\eea
These poles result, respectively, from the zeros of $\cn$, and $\dn$
in the fundamental domain. Each one of the above points $z$
corresponds to a single term in $G$  being singular. Consider now the
behavior of $G$ near the poles of the $B_2$-terms, $z _\pm = K \pm iK'$.
Here, we have
\bea
 \sn (z _\pm, k) & = &  1/k
\no \\
\cn (z _\pm, k) & = &  \mp i k'/k
\no \\
\dn (z _\pm,k) & = & 0
\eea
where $k, k'>0$, and $k^2 + (k')^2=1$. The term in $G$ proportional to
$B_2^*$ is the only one exhibiting a singularity at $z = z _\pm$; the term
proportional to $B_2$ vanishes there, while the terms in $B_1$ and $B_1^*$
are regular and non-zero. The singular and leading finite contributions to $G$
are then given as follows,
\bea
G = \pm { i \over k'} \Big ( B_1 ( 1-k) -  B_1^* (1+k ) \Big )
\pm { i  B_2^*  \over k'}  {1 \over \bar z - \bar z _\pm}
+ \cO \left ( |z -  z _\pm | \right )
\eea
where we have used $\dn (\bar z _\pm) = \mp i k'$. If $0 < |B_2| \ll |B_1|$,
this form of $G$ will always have a zero $z _\pm ^0 $  in the vicinity
of $z _\pm $,  for which we may write down an explicit formula,
\bea
z _\pm ^0 = z _\pm
+ { 2 B_2 \over B_1 (1+k) - B_1^* (1-k)} + \cO \left ( |B_2|^2 \right )
\eea
The implicit function theorem guarantees that such a solution will
exist at least in an open neighborhood of $B_2=0$. (For $B_2=0$, the
pole would be at $z _\pm$, but its residue now vanishes.)

\sm

This result demonstrates that there cannot be a smooth deformation
away from the M-Janus solution, by a ``complex conjugate pair of
eigenvalue solution" which satisfies the differential equation for $G$,
as well as the quadratic constraint $|G| \geq 1$.

\subsection{Real and complex conjugate pair eigenvalues}

The one logical possibility still left open by the above negative results
is the superposition of the $\Phi$-field of the M-Janus solution
(with a pair of complex conjugate eigenvalues of $D^{-1}H$), with
the $\Phi$-field associated with two real eigenvalues. We shall continue
to take the complex eigenvalues to be at $\k = \pm i$, and
denote the real eigenvalues $\pm 1/a$, for $a$ real and positive.
The corresponding $\Phi$ is then given as follows,
\bea
\Phi & = &
{A _+ \over a} (1 + a \k)^\half (1+ a \bar \k )^\half
+ {A _- \over a} (1 - a \k)^\half (1- a \bar \k )^\half
\no \\ &&
+ B (1 + i \k )^\half (1+ i \bar \k )^\half
+ B^* (1 - i \k )^\half (1- i \bar \k )^\half
\eea
In $\k$-coordinates, the function $G$ is given by
$G = 2 i \p_{\bar \k } \Phi$, and we find,
\bea
G = i  A _+ \left ( { 1 + a \k \over 1 + a \bar \k } \right )^\half
- i  A _- \left ( { 1 - a \k \over 1 - a \bar \k } \right )^\half
- B  \left ( { 1 + i \k \over 1 + i \bar \k } \right )^\half
+ B^*  \left ( { 1 - i \k \over 1 - i \bar \k } \right )^\half
\eea
The branch points on $\bR$ at $\pm a^{-1}$ are responsible
for flipping the signs of the corresponding terms there.

\sm

To investigate suitable boundary conditions on $G$ requires disentangling
the branch cut in the complex plane. The branch points at $\k = \pm a^{-1}$
also produce branch cuts, but these could consistently be taken away from
the upper half-plane (as was done when we solved the $AdS_7 \times S^4$
deformations). To this end, we carry out a (partial) uniformization,
which unfolds the branch cuts starting at $\k = \pm i$, and
introduce the variable $u$, related to $\k$ by,
\bea
\k (u) = { 2 u \over u^2-1}
\eea
We use the M-Janus parameterization $B = (i+\lambda)/2$, and uniformize
the square roots in the parameter $a$ by setting,
\bea
a = \sh (\f ) \hskip 0.7in \f >0
\eea
The relevant quadratic functions in $u$ then factorize as follows,
\bea
u^2 \pm 2 a u -1 = \left ( u + e^{\pm \f} \right ) \left ( u - e^{\mp \f} \right )
\eea
These points are ordered as follows,
\bea
- e^\f < -1 < - e^{- \f} < e^{- \f} < +1 < e^\f
\eea
The resulting $G$-functions is then given by,
\bea
G & = & i { \bu ^2 -1 \over |u^2-1|} \left  [
A _+ \left ( { (u+e^\f)(u- e^{-\f} ) \over (\bu + e^\f ) (\bu - e^{- \f}) } \right )^\half
-A _- \left ( { (u- e^\f)(u+ e^{-\f} ) \over (\bu - e^\f ) (\bu + e^{- \f}) } \right )^\half
\right .
\no \\ &&  \hskip 1in \left .
- {  |u|^2 +1 -  \lambda (u - \bu) \over \bu^2+1} \right ]
\eea
The boundary of $\Sigma$ is to be at $\Im (\k) =0$, which
corresponds precisely to $\Im (u)=0$. As $u$ proceeds along the real
axis, the third term in the brackets is constant, and equal to $-1$.
In the first two terms, the functions are piece-wise constant, but
change sign whenever a branch cut is being traversed. The basic
function to be considered is
\bea
f(u) = \left ( { u - u_0 \over \bu - u_0} \right )^\half
\eea
where $u_0$ is real. Assuming that the sign of the square root is
taken to be + for $u > u_0$, we determined the sign for $u <u_0$
by introducing an adapted local coordinate, $u = u_0 +v^2$, in which
we have $f(u) = v / \bar v$.
As we proceed from $u >u_0$ for which $v$ is real so that $f(u)=+1$,
to $u<u_0$ for which $v$ is imaginary, we see that we must have $f(u)=-1$.
Following this procedure through all branch cuts, we have the following
table of boundary values of $G$,
\bea
e^\f < u ~~~~~ & \hskip 1in & G = +i (A _+ - A _- -1)
\no \\
+1 <  u < e^\f  & \hskip 1in & G = +i (A _+ + A _- -1)
\no \\
e^{-\f} < u < +1 & \hskip 1in & G = - i (A _+ + A _- -1)
\no \\
- e^{-\f} < u < e^{- \f} & \hskip 1in & G = - i (- A _+ + A _- -1)
\no \\
-1 < u < - e^{-\f} & \hskip 1in & G = - i (- A _+ - A _- -1)
\no \\
- e^\f < u < -1 & \hskip 1in & G = +i (- A _+ - A _- -1)
\no \\
 u < - e^\f ~~~~~ & \hskip 1in & G = +i (A _+ - A _- -1)
\eea
The overall sign reversals for $-1<u<+1$ are due to the prefactor
$(\bu^2-1)/|u^2-1|$ in $G$. The boundary conditions require that
$G= \pm i$ throughout the real line, or equivalently,
\bea
(A _+ \pm A _- -1)^2 = (- A _+ \pm A _- -1)^2 =1
\eea
where the above $\pm$ signs are uncorrelated. These
conditions require that $A _+ (\pm A _- -1)=0$.
Taking the sum of these two relations, we find
$A _+=0$, and thus $A _-=0$. Thus, one
cannot add the real eigenvalue contributions without violating
the boundary conditions on $G$.

\subsection{Generalization to arbitrary numbers of poles}

The arguments presented in the preceding sections were for 
$p=2$ and $q=1$, but may be generalized to arbitrary $p,q$,
by considering a solution to the differential equation for $G$ 
which has $q$ complex conjugate pairs of eigenvalues. 
For simplicity, we shall limit the discussion here to the case $p=0$.
The expression for $G$ is readily written down in terms
of the holomorphic function $\k$, which  runs through the
upper half-plane, $\Im (\k) >0$,
\bea
G = \sum _{n=1}^N \left [
- B_n \left ( { \k - w_n \over \bar \k - w_n} \right )^\half
+ B_n ^* \left ( { \k - \bar w_n \over \bar \k - \bar w_n} \right )^\half
\right ]
\hskip 0.6in \Im (w_n) >0
\eea
The generalization of the preceding result, which had been obtained for the 
elliptic case only, may be stated as follows. 

\sm

{\sl For sufficiently small $B_m$ the singularity at 
$\k = w_m$ is  a pole which produces a zero of $G$ in its neighborhood.} 

\sm

To prove this, we begin by uniformizing the square root around $w_m$,
in a neighborhood of $w_m$, and introduce a suitable local coordinate $u$,
\bea
\k - w_m = u^2
\eea
The singular and leading finite contributions to $G$ are then given
as follows,
\bea
G & = & G_m + B_m^* { \sqrt{ w_m - \bar w_m} \over \bar u}
\no \\
G_m & = & \sum _{n\not= m} \left [
- B_n \left ( { w_m - w_n \over \bar w_m - w_n} \right )^\half
+ B_n ^* \left ( { w_m - \bar w_n \over \bar w_m - \bar w_n} \right )^\half
\right ]
\eea
For each $m$, the quantity $G_m$ is a finite complex number.
If $G_m$ happens to vanish (which is not generic), then $B_m=0$
produces a zero of $G$. If $G_m \not= 0$, then $G$ has a
zero in the neighborhood of $w_m$, for which we may give an
explicit formula,
\bea
u = B_m { \sqrt{\bar w_m - w_m} \over G_m^*} + \cO \Big ( |B_m|^2 \Big )
\eea
Thus, we conclude that, within the approximation of small deviations
away from the maximally symmetric solution, the bound $|G|>1$
is always violated by a region containing a point where $G=0$.

\newpage

\section{M-brane interpretation of rigidity}
\setcounter{equation}{0}

There are several arguments in M-theory for the existence of  half-BPS brane 
configurations whose geometry is $AdS_3 \times S^3 \times S^3$ warped over a 
two-dimensional surface $\Sigma$. We list these arguments below, and discuss 
their various qualifications and limitations.
\begin{itemize} 
\item  
The first argument originates from considering the symmetries of probe branes, 
such as M2 probe branes inserted into $AdS_7 \times S^4$, or M5 probe branes inserted 
into $AdS_4 \times S^7$. The embedding of the probe branes preserves the same 
symmetries as our Ansatz for full solutions does. The preservation of half of the supersymmetries of the probe branes is assured, in the Green-Schwarz formalism, 
by the existence of a kappa-symmetry projector in the world-volume theory of the 
probe brane \cite{Skenderis:2002vf}. While the presence of probe branes is 
suggestive of the existence of back-reacted supergravity solutions, only the resolution of
the full supergravity BPS equations (as is being analyzed here) can conclusively 
decide on the existence of any full fledged regular supergravity solutions.
\item 
The second argument rests on the existence of asymptotic symmetries
and supersymmetries  in AdS space-times. The existence of full fledged
half-BPS supergravity solutions with local $AdS_4 \times S^7$ asymptotics 
requires the existence of certain subalgebras with 16 fermionic generators
of $OSp(8|4,\bR)$. These subalgebras do indeed exist \cite{D'Hoker:2008ix}.
Conversely, however, the mere existence of a corresponding subalgebra 
does not suffice to guarantee existence of a full-fledged regular supergravity solution.
\item 
The third argument is based on the expected   symmetry enhancement, 
in the near-horizon limit, for the 1+1 dimensional intersection of M2- and M5-branes.
The full regular supergravity solution with M2- and M5-branes intersecting over a 
1+1-dimensional space-time is expected to exist and preserve 1/4 of the 
supersymmetries of flat space-time. Enhancement in the near-horizon limit would 
then result in a regular half-BPS supergravity  solution. 
Note, however, that no such localized intersecting brane solutions are known
explicitly at this time, so that we do not know for sure which near-horizon 
limits exist, and which do not. Thus, the third argument proceeded mostly by 
analogy with known cases in Type IIB, but does not guarantee existence 
of full fledged regular supergravity solutions either.
\end{itemize}

All three arguments given above would proceed in parallel for the cases
of solutions asymptotic either to $AdS_4 \times S^7$ or to $AdS_7 \times S^4$.
The arguments would suggest the existence of full fledged half-BPS solutions
asymptotic to  $AdS_4 \times S^7$ just as they do for solutions
asymptotic to $AdS_7 \times S^4$.
Having critically discussed the various arguments above {\sl in favor} of the 
existence of regular supergravity solutions with $AdS_4 \times S^7$ asymptotics,
we shall now refine the arguments by exhibiting a fundamental asymmetry
between the case of solutions asymptotic to $AdS_4 \times S^7$ and the 
case asymptotic to $AdS_7 \times S^4$. This asymmetry will be key in our
final argument against the existence of regular half-BPS supergravity solutions 
which are asymptotic to $AdS_4 \times S^7$ and carry M5-brane charges.

\subsection{Asymmetry between M2- and M5- branes}

The conserved charges associated with M5- and M2-branes are given by,
\be
\label{mcharge}
Q_{5}= \int_{{\cal M}_{4}} F_{4},  \hskip 1in  Q_{2}
= \int_{{\cal M}_{7}} (*F_{4}+  F_{4} \wedge C_{3})
\ee
where ${\cal M}_{4}$ and ${\cal M}_{7}$ are surfaces surrounding the M5- and M2-brane, respectively, capturing all the flux. From the arguments given at the beginning 
of this section one would naively expect that there exist two kinds of  half-BPS 
solutions: First,  fully back-reacted solutions asymptotic to $AdS_{7}\times S^{4}$ 
which also support M2-brane charge and second,  fully back-reacted solutions 
asymptotic to $AdS_{4}\times S^{7}$ which also support M5-brane charge.

\sm

Solutions corresponding to the first case do indeed exist and were explicitly 
constructed in  \cite{D'Hoker:2008qm}. These solutions have one asymptotic 
$AdS_{7}\times S^{4}$ region and in addition non-trivial seven cycles with 
non-zero $Q_{2}$.

\sm

The arguments and calculations presented in the present paper show, 
however, that the corresponding solutions for the second case do not exist. 
In other words, even though  the general fully back-reacted  half-BPS solution 
which is asymptotic to  $AdS_4 \times S^7$\ has non-trivial 4-cycles, 
they do not support any  net M5-brane charge. In fact, the only new half-BPS 
solution found here and in \cite{D'Hoker:2009gg} is the M-Janus solution
and its multiple covers, but these solutions have no net M5-brane charges.

\sm

The asymmetry of the two cases may seem surprising at first , given the fact that 
the local solutions are very similar. In the following we give an explanation for 
the different behavior of the two cases. Two key Theorems enter into consideration,

\sm

\noindent 
{\bf Theorem 1}

{\sl M2-branes can end on M5-branes, but M5-branes cannot end on
M2-branes.}

\medskip

\noindent
{\bf Theorem 2} 

{\sl No solutions exist which have $SO(2,2) \times SO(4) \times SO(4)$ symmetry
and 16 supersymmetries throughout space-time and possess both 
$AdS_4 \times S^7$ and $AdS_7 \times S^4$ asymptotic regions.}

\medskip

\noindent
The result of Theorem 1 dates back to the early days of M-theory (see for example
\cite{Strominger:1995ac,Townsend:1996em}), and was obtained using the 
Chern-Simons term in M-theory.
The argument for why  an  M5-brane can not  end on an M2-brane is as follows. 
The 4-surface  $\cM_{4}$ in the formula (\ref{mcharge})  for the M5-brane 
charge surrounds the M5-brane transversely to capture all the flux. If the 
M5-brane ends on an M2-brane, this surface can be slipped off and contracted  
and we arrive at a contradiction,  since one calculation of the M5-brane charge 
gives a non-zero while the other calculation gives zero.

\begin{figure}\label{figure2}
\centering
\includegraphics[ scale=0.6]{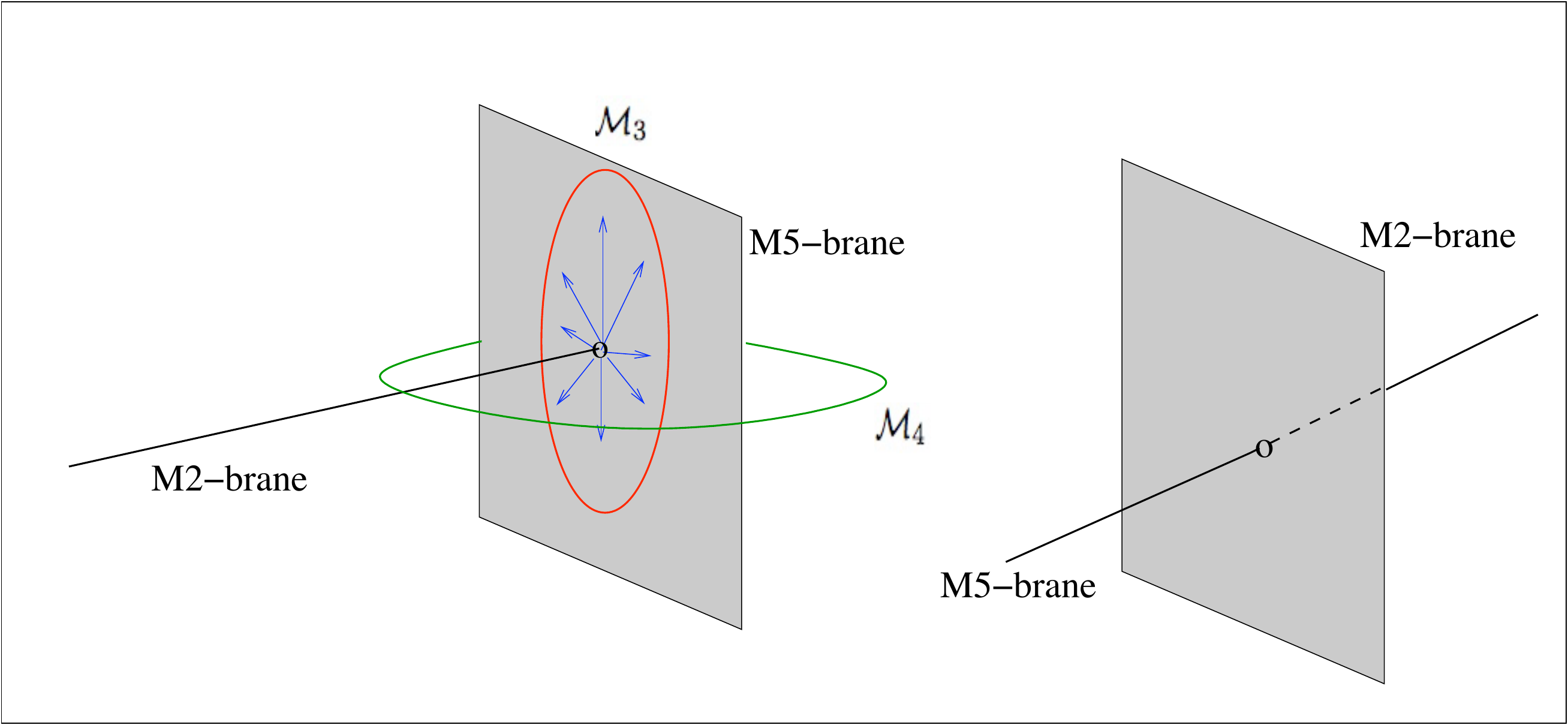}
\caption{An  M2-brane ending on an M5-brane produces flux inside the 
M5-brane worldvolume. By contrast, an M5-brane can only intersect an M2 brane
(see also \cite{Townsend:1996em}).}
\end{figure}

%

\noindent
The existence  of a Chern-Simons term in the 11-dimensional supergravity 
action implies the presence of the $F_{4}\wedge C_{3}$ term  in  the formula 
for the M2-brane charge (\ref{mcharge}).  This term makes it possible for an 
M2-brane  to end on an M5-brane without violating charge conservation. 
As one  tries to slip off the seven surface ${\cal M}_{7}$ in (\ref{mcharge}), 
the surface decomposes into a 4-surface ${\cal M}_{4}$ enclosing the 
M5-brane on which the M2-brane ends and a 3-surface ${\cal M}_{3}$ lying in 
the world-volume of the M5-brane enclosing the two dimensional boundary 
of the M2-brane,
\be
Q_{2}= \int_{{\cal M}_{4}} F_{4} \int _{{\cal M}_{3}} C_{3} = Q_{5} \int _{{\cal M}_{3}} C_{3}
\ee
An M2-brane ending on a M5-brane induces a non-trivial 3-form 
field $C_{3}$ in the world-volume of the M5-brane. This supergravity field is 
related to the self-dual 3-form field strength of the world-volume theory of 
the M5-brane \cite{Howe:1997fb,Aganagic:1997zq,Bandos:1997ui}.

\sm

The result of Theorem 2 was obtained in \cite{D'Hoker:2008wc}, and derived from 
the explicit construction of the local solutions.  The general half-BPS 
solution is characterized by three real constants $c_{i}, i=1,2,3$, which are subject 
to the condition $c_{1}+c_{2}+c_{3}=0$.
The constants $c_{i}, i=1,2,3$ are related to the normalization of the metrics of the 
$AdS_3 \times S^3 \times S^3$ factors respectively. An overall scaling of all $c_{i}$ 
can be absorbed into a rescaling of the eleven dimensional metric and is not important. 
The key fact is that the choice of $c_{i}$ characterizes the asymptotic geometry of the  
half-BPS solution completely. In particular  the choices $c_{1}=c_{2}$ and $c_{1}=c_{3}$   
imply asymptotic $AdS_{7}\times S^{4}$
regions,  whereas $c_{2}=c_{3}$ implies asymptotic $AdS_{4}\times S^{7}$ regions.  
Because of the  $c_{1}+c_{2}+c_{3}=0$ constraint the conditions can never be  
satisfied simultaneously. It follows that a half BPS-solution cannot have asymptotic 
$AdS_{7}\times S^{4}$   and  $AdS_{4}\times S^{7}$ regions at the same time.

\subsection{Comparison  to solutions asymptotic to $AdS_7 \times S^4$}

The argument presented above shows that if an M2-brane ends on (a stack of) 
M5-branes, the net flux carried by
the M2-brane is conserved and must be carried off by the gauge field
on the M5-brane. No matter which limit we take of the M5-brane (such
as the near-horizon limit), the net flux will continue to produce a
non-zero M2-brane charge on the M5-brane. Therefore, in the
near-horizon limit, we should  expect to recover an $AdS_7 \times S^4$
space-time with net M2-brane charges inserted at the remnant points
due to the M2-branes. This is indeed borne out by the explicit and
exact solutions of \cite{D'Hoker:2008qm}, which exhibit net M2-brane
charges in an asymptotic $AdS_7 \times S^4$ space-time. 

\sm 

Theorem  2 implies that, given the $AdS_7 \times S^4$ asymptotic solution with
non-trivial M2-brane charges, {\sl there should exist no limit in which we
can recover the M2-brane geometry.} Indeed, if such a M2-brane
near-horizon limit existed, then we would have a geometry which
exhibits both an $AdS_4 \times S^7$ asymptotic region (namely near the
M2-brane) and an $AdS_7 \times S^4$ asymptotic region (namely
space-time away from the M2-brane insertions), but this is impossible
in view of Theorem  2. This result is also borne out by the explicit
solution of \cite{D'Hoker:2008qm}. Its only moduli are the positions $x_m$
of the sign-flips in $G$ on the boundary of $\Sigma$. Thus, the only
limits we can consider is to let consecutive positions $x_m$ collapse
to one another. 

\sm

We compute explicitly the M2-charges and show a singular M2-brane
never emerges in the limit of collapsing moduli.
Using the general expression for $G$ in (3.28) of \cite{D'Hoker:2008qm}, 
as well as the expressions for the fluxes in (2.8) of \cite{D'Hoker:2008qm}, 
we may compute the explicit fluxes in an asymptotic expansion away from 
the boundary for the case of general $g$.  The result is
\bea
\p_w b_i = 
-  2i \ep_i -i \eta_i  \sum_{n=1}^{g+1} 
\bigg( {x-x_{2n-1} \over |x-x_{2n-1}|} - {x-x_{2n} \over |x-x_{2n}|} \bigg) + \cO(y) 
\eea
The values of the sign factors $\ep_i$ and $\eta _i$ are given by
$\ep _1=-\ep_2=+1$, $\ep_3=0$, and $\eta _1=\eta_2=-\eta_3=1$, and the 
points are ordered as follows,\footnote{The notation 
in terms of the points $x_n$ used here is related to the notation in terms
of points $a_n, b_n$ of \cite{D'Hoker:2008qm} as follows:  $a_n = x_{2n}$, 
and $b_n = x_{2n-1}$ for $n=1, \cdots, g+1$.}
\bea
x_1 < x_2 < x_3 <  \cdots < x_{2g} < x_{2g+1} < x_{2g+2}
\eea
In general, the geometry contains $2g+1$ four-cycles, which split into two groups.  
One group contains $g$ four-cycles formed from the first three sphere $S_2^3$, 
while the second group contains $g+1$ four-cycles formed from the second  
sphere $S_3^3$.  The charge along the first group can be computed by integrating 
the imaginary part of $\p_w b_2$ along the boundary starting from $\xi_n$ and 
ending at $\xi_{n+1}$ with $n = 1,...,g$ and $x_{2n-1} < \xi_n < x_{2n}$.
The charge along the second group of four cycles can be computed by integrating 
the imaginary part of $\p_w b_3$ along the boundary starting from $\xi'_n$ and 
ending at $\xi'_{n+1}$ with $n = 0,...,g$ and $x_{2n} < \xi'_n < x_{2n+1}$.  
Here we take $x_0 = -\infty$ and $x_{2g+3} = +\infty$.  The resulting charges 
are independent of the choices for $\xi_n$ and $\xi'_n$ and are given as
\bea
Q_{5n}^{(2)} & = & q_5^{(2)} (x_{2n+1} - x_{2n}) \qquad n=1,...,g 
\no \\
Q_{5n}^{(3)} & = & q_5^{(3)} (x_{2n} - x_{2n-1}) \qquad n=1,...,g+1
\eea
where $Q_{5n}^{(2)}$ is the charge along $S^3_2$ and $Q_{5n}^{(3)}$ is the 
charge along $S^3_3$. Here,  $q_5^{(2)}$ and $q_5^{(3)}$ are constants, i.e.
parameters which are independent of the points $x_{n}$. Taking 
$x_{2n+1} \rightarrow x_{2n}$ we see that $Q_{5n}^{(2)} \rightarrow 0$ while 
taking $x_{2n} \rightarrow x_{2n-1}$ we see that $Q_{5n}^{(3)} \rightarrow 0$.  

\sm

To compute the M2-brane charge, we first note that the integral of $*F_{4}$ over 
the seven cycle formed by the product of one of the four-cycles and the conjugate 
three sphere always vanishes.  This is because we must integrate the real part of 
$\p_w b_1$ along the boundary which always vanishes.  This shows that the 
M2-brane charge comes purely from the topological term and takes the value
\bea
Q_{2i}^{(2)} = \int_{S^4_{2i}} F_{4} \int _{S^3_3} C_{3} = Q_{5i}^{(2)} \int _{S^3_3} C_{3}
\no\\ 
Q_{2i}^{(3)} = \int_{S^4_{3i}} F_{4} \int _{S^3_2} C_{3} = Q_{5i}^{(3)} \int _{S^3_2} C_{3}
\eea
Collapsing branch cuts, we see that the vanishing of the M5-charge 
immediately implies the vanishing of the M2-charge.  Thus, in the limit 
consecutive positions $x_m$ collapse to one another we see that the 
net M2-brane charge localized at these points always tends to 0.

\subsection{Completing arguments for solutions asymptotic to $AdS_4 \times S^7$}

The arguments for the absence of half-BPS solutions which are asymptotic to 
$AdS_{4}\times S^{7}$ and support M5-brane charge can now be completed. 

\sm

Theorem  1,  given above,  asserts that an M5-brane cannot end on an 
M2-brane. 
Hence there is no net 4-form flux to fan out into the M2-brane geometry that produces 
a non-vanishing net M5-brane charge.  Of course, the M5-brane can simply intersect 
the M2-brane, but this configuration produces no net  charge into the M2-brane. 
In the near-horizon limit, no M5-brane flux is required to survive.

\sm

Theorem  2,  given above,  implies  that  there exists
no near-horizon limit in which both the M2-brane and the M5-brane
survive (since this limit would exhibit both $AdS_4 \times S^7$ and
$AdS_7 \times S^4$ asymptotic regions). Hence, the near-horizon
limit will be either that of an M2-brane  or of an M5-brane, but not of both 
simultaneously.  This implies that the  $AdS_4 \times S^7$
asymptotic space-time  cannot sustain any conserved 5-brane charge.

\sm

These arguments seem to be borne out by the fact that on the one hand the 
explicit solutions  we have found in this paper, i.e. the multi cover M-Janus 
solutions of section 3, have four cycles and a nontrivial 4-form field, but that 
on the other hand the integrated M5-brane charge along these cycles vanishes. 
We discuss possibles loopholes in this result in the next section.

\newpage

\section{Discussion}
\setcounter{equation}{0}

The main result of the present  paper is that regular half BPS-solutions of 
M-theory that enjoy $SO(2,2)\times SO(4)\times SO(4)$  symmetry and  
are asymptotic to  $AdS_{4}\times S^{7}$ are remarkably rigid in the sense 
that the only non-trivial solutions we have found is the M-Janus solution of  
\cite{D'Hoker:2009gg} and multiple covers thereof.  We have given an 
interpretation of this result in terms of M2- and M5-brane intersections
and endings in the previous section. In this section, we shall conclude
with a discussion of a number of questions left open by our work, and of 
possible directions for further research.

\subsection{Origin and physical significance of multiple covers of M-Janus}

The first open question concerns the physical interpretation of the multiple 
covers of the M-Janus solution found in section \ref{secthree}. 
The M-Janus solution  \cite{D'Hoker:2009gg} has two asymptotic  $AdS$ 
regions. Its interpretation in the dual 2+1-dimensional CFT  is  given by
the insertion of a dimension 2 operator  localized along a 
1+1-dimensional linear interface/defect, thereby partially breaking the 
superconformal symmetry. The $(g+1)$-fold cover of the M-Janus solution in turn  
has $2g+2$  asymptotic $AdS$ regions.  The behavior of the supergravity 
fields is identical to that of the parent M-Janus solution with only two asymptotic 
regions, and can in principle be smoothly projected to the parent M-Janus solution 
by a freely acting discrete transformation group. Furthermore, 
the solution does not support any M5-brane charge.

\sm

It is useful to contrast this solution with the corresponding multi-Janus solution 
of Type IIB supergravity which was constructed in \cite{D'Hoker:2007xz}. The 
Type IIB multi-Janus solution also has  $2g+2$  asymptotic $AdS$ regions. 
However, in the Type IIB solution,   the dilaton generically takes different   
values in the different  asymptotic regions. In the dual CFT this means that the 
theories on the different half-spaces, which are glued together at the defect, 
are all different.   In addition, the solution supports non-zero five brane charge.
In M-theory, the bulk CFTs dual to each asymptotic region have no free 
parameters, since the gauge coupling of the maximally supersymmetric 
2+1-dimensional CFT is fixed. This may be the source of the identical nature
of the cover copies in the corresponding supergravity solutions.

\sm

It would be interesting to investigate  the multiple cover solutions further. 
On the supergravity side it is possible to calculate correlation functions of 
operators located in different asymptotic regions, and elucidate whether
the multiple-cover M-Janus solutions are genuinely different from the 
parent solutions with only two asymptotic $AdS$ regions. It would also be interesting 
to determine whether a sensible CFT interpretation of the multiple defect 
theories exists.  

\sm

A great deal of progress has been made in understanding 
the CFT associated with the decoupling limit of  multiple M2-branes 
\cite{Bagger:2006sk,Bagger:2007jr,Gustavsson:2007vu}. Of particular 
interest is the ABJM solution of \cite{Aharony:2008ug} which has manifest $\cN=6$ 
supersymmetry and is dual to M-theory on the quotient 
$AdS_{4}\times S^{7}/Z_{k}$. The 
M-Janus solution admits a regular ABJM reduction to a quotient solution which is
invariant under $SO(2,2) \times SO(4)\times U(1)^2$, preserves 12
supersymmetries, and provides a Janus-like interface/defect  solution in
ABJM theory.  It would be interesting to analyze the multiple cover M-Janus solution 
and its dual CFT interpretation in the context of the ABJM quotient as well.

\subsection{Can one relax the regularity and boundary conditions ?}

The second open question concerns the assumptions we have made on
the  boundary and regularity conditions. 
The rigidity result we  found is closely related to the regularity 
assumptions and the constraint $|G|^{2}>1$, which the solution 
has to satisfy point-wise. It is easy to construct more general solutions which 
satisfy the differential equation (\ref{G1}) by linear superposition.
In all the cases we have explored, however,  
the solution has one or more of the following properties: 
\begin{enumerate}
\item the solution is singular;
\item  the solution is not asymptotic to $AdS_{4}\times S^{7}$;
\item the solution violates the constraint $|G|^2>1$.
\end{enumerate}
It is an open question wether it is possible to modify or omit some of the 
assumptions in order to obtain new solutions which have a sensible physical 
interpretation. Below, we shall list three scenarios in which weaker conditions
appear.

\sm

$\bullet$ First, there exists an example of such a solution in the literature \cite{Boonstra:1998yu,Gauntlett:1998kc,deBoer:1999rh} which violates assumption 2.  It is characterized 
by a space-time manifold $AdS_{3} \times S^{3} \times S^{3} \times T^2$ 
without any warping. The solution is regular, but not asymptotic to 
$AdS_{4}\times S^{7}$.   A systematic analysis of  more general solutions with 
other asymptotics would be interesting, but is beyond the scope of this paper.

\sm

$\bullet$ Second, the absence of any M5-brane charge in our regular solutions 
may suggest that if one drops the condition of regularity on the boundary 
$\p\Sigma$, it  might be possible  to obtain solutions which have non vanishing 
M5-brane charge. A particular generalization is  obtained  by choosing a 
harmonic function $h$ of the type given in (\ref{h}), but which contains higher order poles. 
Another possibility is to  allow for isolated points  on the inside of $\Sigma$ where the 
constraint $|G|^{2}>1$ is violated, and we have instead $|G|^2=1$. It is an 
open question whether such  solutions  are sensible. One possibility is that 
singular solutions may be associated with the presence of probe M5-branes 
in the $AdS_{4}\times S^{7}$ spacetime.

\sm

$\bullet$ Last, it might be the case that demanding the preservation of 16 supersymmetries is too restrictive to allow for regular solutions. A way out 
might be to look for solutions which only preserve 8 supersymmetries, yielding
quarter-BPS solutions.  Intersecting M2/M5-brane solutions in flat space preserve 
one quarter of the 32 Minkowski supersymmetries  and it is possible that the 
near-horizon limit does not lead to an enhancement of the number of 
supersymmetries from 8 to 16.  We leave these interesting questions for  future work.

\bigskip\bigskip

\noindent{\Large \bf Acknowledgements}

\medskip

MG gratefully acknowledges the hospitality of   the Department of Physics and 
Astronomy, Johns Hopkins University, during the course of this work.

\newpage

\appendix

\section{Reduced form of the polynomial $P$}
\setcounter{equation}{0}

The form of the polynomial $P$ in the Ansatz of (3.16) does not account 
for the most general real polynomial of degree 4. In this appendix, we shall
derive the general restrictions on $P$ of (3.9) that result from the 
polynomial equations (3.10) and the boundary conditions.
To construct explicit solutions, and describe their moduli spaces concretely,
it will be helpful to reduce the Ansatz of (\ref{phans}) by demonstrating that
$P$ is not a general polynomial of degree $2g+2$ in $u,\bu$, but obeys
certain restrictions, as a consequence of the boundary conditions and
of the polynomial equation (\ref{polyrel}).

\sm

$ \bullet$ The {\sl boundary conditions} force $P$ to consist of an arbitrary 
polynomial of
degree $2g+1$ plus a single term of the form $(u \bar u)^{g+1}$.  To obtain
this result, we note that $u \to \infty$ is part of the boundary of $\Sigma$ on which
we must have $|G|= 1$.  The magnitude of $G$ may be readily evaluated from (\ref{G1}).
The asymptotic behavior as $u \to \infty$ of the denominator of $G$ is,
\bea
\bar Q \p_{\bar u} \bar R - \bar R \p_{\bar u} \bar Q = (2g+1) \bar u^{4g+2}
\sum_{i=1}^{2g+2} c_i
\eea
Considering now the behavior of a term $P_{mn}=u^m \bar u^n \pm u^n \bu ^m$ in $P$
as $u \rightarrow \infty$,  we find,
\bea
P_{mn} \p_{\bar u} \bar Q - 2 \bar Q \p_{\bar u} P_{mn}= (2g+2-2n)  u^m \bar u^{n+2g+1} \pm
(2g+2-2m)  u^n \bar u^{m+2g+1}
\eea
Requiring the ratio of the numerator and the denominator of the expression
for $|G|$ to be finite as $u \to \infty$ yields the following allowed orders $m,n$,
\bea
m=n & = & g+1
\no \\
m + n & \leq & 2g+1
\eea
which proves our above assertion.

\sm

$\bullet$ The {\sl polynomial relation} (\ref{polyrel}) will schematically be 
represented by $\cD P=0$. This relation forces further restrictions. It will be 
instructive to derive these  independently
of the restrictions derived above from the boundary conditions.
The most general real polynomial $P$ of degree $2g+2$ in $u,\bu$, may be
parametrized as follows,
\bea
\label{phansb}
P(u, \bar u) = \sum _{m=0}^{2g+2}  \sum _{{0\leq n \leq m \atop m+n \leq 2g+2} }
\! \! \! d_{m,n} (u^m \bar u ^n + u^n \bar u^m)
+ \sum _{m=0}^{2g+2} \sum _{{0\leq n \leq m-1 \atop m+n \leq 2g+2} }
\! \! \! i e_{m,n} (u^m \bar u ^n - u^n \bar u^m)
\eea
The polynomial $P$ is real provided $d_{m,n}$ and $e_{m,n}$ are real.
The number of free parameters for $d_{m,n}$  is $(g+2)(2g+3)$, while for
$e_{m,n}$, it is $(g+1)(2g+3)$, giving a total of $(2g+3)^2$.
Parts of the polynomial relation $\cD P(u,\bu)=0$ may be solved by an iterative process.

\sm

The following restrictions are found to arise,
\begin{enumerate}
\item All coefficients $d_{m,n}$ with $m > g+1$ must vanish;
\item All coefficients $e_{m,n}$ must vanish.
\end{enumerate}
These results may be shown iteratively beginning with the coefficients
corresponding to the highest value of $m-n$, which are $d_{2g+2,0}$
and $e_{2g+2,0}$. Equations in which these coefficients appear as overall factors,
show that these coefficient must vanish.
Setting now $d_{2g+2,0}=e_{2g+2,0}=0$, one finds that there are factorized
equations also for $d_{2g,2}$ and $e_{2g,2}$, so that also these coefficients
must vanish and so on. In this manner, one shows iteratively that $d_{m,n}=e_{m,n}=0$
for all $m > g+1$. (Note that the coefficients $d_{m,n}$ and $e_{m,n}$ are
defined with $m \geq n$, so that  $n > g+1$ automatically implies $m > g+1$,
and these cases does not have to be considered separately.)

\sm

To show that the coefficients $e_{m,n}$ for the remaining range of $m,n$
also vanish, one proceeds to split the polynomial relation $\cD P(u,\bu)$
into real and imaginary parts. These are correlated with the appearance
of the coefficients $d_{m,n}$ and $e_{m,n}$, as follows,
\bea
(\cD P) (u,\bu) - (\cD P)(\bu, u) & ~~ \hbox{contains only} & d_{m,n}
\no \\
(\cD P) (u,\bu) + (\cD P)(\bu, u) & ~~ \hbox{contains only} & e_{m,n}
\eea
The vanishing of all contributions containing only $e_{m,n}$ may be
solved iteratively again, each time encountering at least one coefficient
$e_{m,n}$ in factorized form. We have checked these results explicitly
by MAPLE for $g =0,1,2,3,4$.

\sm

In summary, incorporating the above results, the form of the polynomial
Ansatz is drastically simplified, and we are left with the following remaining form,
\bea
\label{Phi2}
P(u, \bar u) = \sum _{m=0}^{g+1} \, \sum _{n=0} ^m
d_{m,n} \left ( u^m \bar u ^n + u^n \bar u^m \right )
\eea
Note that the term of maximal degree is unique, and given by $2d_{m, m} (u\bu)^m$
with $m=g+1$. As a result, the point $u = \infty$ is a regular point of the functions
$\Phi$ and $G$.

\newpage

\section{Derivation of the complex analytic equations}
\setcounter{equation}{0}

In this section, we shall show that the assumption of Hermitian pairing in (4.1)
together with the boundary conditions of (4.2) leads to a Cauchy-Riemann
type differential equation for the functions $\phi_m$, given in (4.3).
We shall now proceed to reducing the partial differential equations
to a set of differential equations in just the holomorphic coordinate $u$. This
reduction is general, and will be proven with the help of a Lemma below.

\medskip

\noindent
{\bf \large Lemma 1}

{\sl Let $\xi_m(u)$ and $\zeta _m(u)$ with $m=1, \cdots, N$ be meromorphic functions
of $u$, and let the set $\{ \xi _m (u)\}$ consist of $N$ linearly independent  functions.
Both $\xi_n(u)$ and $\zeta _m(u)$ are assumed to be real, in the sense that
$\overline{\xi_m (u)} = \xi _m (\bu)$ and $\overline{\zeta _m (u)}= \zeta _m (\bu)$.
If the following pairing
\bea
\sum _{m=1}^N \overline{\zeta _m (u)} \xi _m (u)
\eea
is real for all complex $u$, then there exists a real symmetric $N\times N$ matrix $H$,
whose coefficients are independent of $u, \bu$, such that
\bea
\zeta _m (u) = \sum _{n=1}^N H_{mn} \xi _n(u)
\eea}
To prove the Lemma, we make use of the reality of the functions to express the
complex conjugates $\overline{\zeta _m(u)}$ of the function $\zeta _m(u)$ in terms
of the same function $\zeta _m(v)$ evaluated on the complex conjugate variable
$v=\bu$ instead. The reality condition of the pairing may then be written as follows,
\bea
\sum _{m=1}^N \Big ( \zeta _m (v) \xi _m (u) - \zeta _m (u) \xi _m (v) \Big )=0
\eea
As usual, we may now view $v$ as a complex variable which is completely
independent of $u$. We may now regard the functions $\xi_m(u)$ as a given
set of linearly independent meromorphic functions, and seek to establish the
relation of the $\zeta _m$-functions with $\xi_m$. The linear independence
of the functions $\xi _m(u)$ for $m=1, \cdots ,  N$ guarantees that we can
pick $N$ points $a_n$ with $n=1,\cdots, N$ such that the $N \times N$
matrix $\xi _m (a_n)$ is invertible. We denote its inverse by $X_{mn}$,
so that
\bea
\sum _{p=1}^ N \xi _m (a_p)  X_{pn}  = \delta _{mn}
\eea
Hence, we have
\bea
\zeta _m (u) = \sum _{n=1}^N  H_{mn} \xi _n (u)
\hskip 1in
H_{mn} = \sum _{p=1}^N \zeta _n (a_p) X_{pm}
\eea
The matrix $H$ is independent of $u$ by construction. It is also independent
of the choice of points $a_p$. A non-zero meromorphic function cannot vanish
identically on the real axis. Therefore, the points $a_p$ may be chosen on the
real axis, and thus $H_{mn}$  is real. That the matrix $H$ is also symmetric
may be established by substituting the relation $\zeta = H \xi$ back into the
equation, which gives,
\bea
\sum _{m,n=1}^N H_{mn} \Big ( \xi_m (u) \xi _n (v) - \xi _m(v) \xi _n(u) \Big )=0
\eea
Only the anti-symmetric part of $H$ contributes in this equation.
If the anti-symmetric part were non-zero, the equation, for any fixed $v$,
would imply a non-trivial  linear relation between the functions $\xi _m(u)$. Since
the $\xi _m$ were assumed to be linearly independent, such a relation must
be trivial, forcing $H_{mn}-H_{nm}=0$.

\end{document}